\documentclass[onecolumn]{aastex62}
\usepackage[encapsulated]{CJK}
\usepackage{mathrsfs}
\usepackage{subfigure}
\usepackage{epstopdf}
\usepackage{amsmath}
\usepackage{longtable}
\usepackage{booktabs}
\usepackage{hyperref} 
\usepackage{graphicx} 
\usepackage{overpic}
\def\etal {et al.~}

\newbox\grsign \setbox\grsign=\hbox{$>$} \newdimen\grdimen \grdimen=\ht\grsign
\newbox\laxbox \newbox\gaxbox
\setbox\gaxbox=\hbox{\raise.5ex\hbox{$>$}\llap
     {\lower.5ex\hbox{$\sim$}}}\ht1=\grdimen\dp1=0pt
\setbox\laxbox=\hbox{\raise.5ex\hbox{$<$}\llap
     {\lower.5ex\hbox{$\sim$}}}\ht2=\grdimen\dp2=0pt

\shorttitle{The PW relation based on Galactic open cluster Cepheids}
\shortauthors{Lin \etal}

\definecolor{malachite}{rgb}{0.34, 0.7, 0.22}

\begin{document}

\title{Calibrating the Cepheid Period--Wesenheit Relation in the \emph{Gaia} Bands using Galactic Open Cluster Cepheids
}

\correspondingauthor{Ye Xu}
\email{xuye@pmo.ac.cn}

\author{Zehao lin}
\author{Ye Xu}
\author{Chaojie Hao}
\author{Dejian Liu}
\affiliation{Purple Mountain Observatory, Chinese Academy of Sciences, Nanjing 210008, People's Republic of China}
\affiliation{University of Science and Technology of China, 96 Jinzhai Road, Hefei 230026, People's Republic of China}
\author{Yingjie Li}
\affiliation{Purple Mountain Observatory, Chinese Academy of Sciences, Nanjing 210008, People's Republic of China}
\author{Shuaibo Bian}
\affiliation{Purple Mountain Observatory, Chinese Academy of Sciences, Nanjing 210008, People's Republic of China}
\affiliation{University of Science and Technology of China, 96 Jinzhai Road, Hefei 230026, People's Republic of China} 

\begin{abstract}
Establishing the period--Wesenheit relation requires independent and accurate distance measurements of classical Cepheids (DCEPs). The precise distance provided by an associated open cluster independently calibrates the period--Wesenheit relation of DCEPs. 51 DCEPs associated with open clusters are compiled using the constraints of five-dimensional astrometric information. {By directly using \emph{Gaia} DR3 parallax,} the period--Wesenheit relation in the \emph{Gaia} $G$ band is calibrated as $W_G = (-3.06 \pm 0.11) \log P + (-2.96 \pm 0.10)$. Compared with the results derived by directly adopting the DCEP parallaxes or using distance moduli, the Wesenheit magnitudes based on the cluster-averaged parallaxes exhibit a tighter relation with the period. Besides, there is a systematic offset between the observed Wesenheit absolute magnitudes of distant OC-DCEPs and their fitted magnitudes. After considering the parallax zero-point correction, the systematic offset can be reduced, yielding a probably better PW relation of $W_G = (-2.94 \pm 0.12) \log P + (-2.93 \pm 0.11)$.

\end{abstract}

\keywords{stars: distances -- stars: variables: Cepheids -- open clusters and associations: general}

%

\section{Introduction}
\label{intro}

Since the discovery of the well-established period--luminosity (PL) relation \citep[also known as ``Leavitt’s law''][]{Leavitt_Pickering_1912}, classical Cepheids, also known as Delta Cepheids (DCEPs), are widely used as universal distance indicators. Benefiting from their accurate distances, DCEPs have been utilized to research the structure and kinematics of the Milky Way \citep{Skowron+2019,Chen+2019}, measure distances to nearby galaxies \citep{Sandage_Tammann_2006}, and constrain the Hubble constant \citep{Riess+2018,Riess+2021}. Therefore, establishing the PL relations of DCEPs has been a major goal and is of primary importance for the distance scale of the Milky Way and nearby galaxies.
Due to the sparse distribution of DCEPs throughout the Galaxy, different lines of sight and distances will produce various reddening effects. Each DCEP has a unique reddening value and is attenuated by intervening interstellar dust to varying extents \citep{Madore+2017}. To reduce the influence of extinction, multiband photometry is usually employed to obtain reddening-free magnitudes, such as the widely used Wesenheit magnitude \citep{Madore_1982,Majaess+2008,Wang+2018}. When adopting a constant extinction ratio, $R_V$, the Wesenheit magnitude can be used directly for DCEP distance measurements as a reddening-independent period--Wesenheit (PW) relation.

Using the trigonometrical parallaxes of DCEPs provided by the Hipparcos mission \citep{ESA+1997}, \cite{Feast_Catchpole_1997} measured the zero-points of the PL and PW relations using the Cepheid distance scale. With the data release of the high-precision astrometric \emph{Gaia} satellite \citep{Gaia+2016,Gaiadr2+2018,Gaiaedr3+2021,Gaiadr3+2022}, a lot of work has been done to calibrate the PL and PW relations using individual DCEPs \citep{Ripepi+2019,Ripepi+2022,Ripepi+2022b,Riess+2021,Owens+2022}. By directly using trigonometrical parallaxes of 898 DCEPs and applying a zero-point correction \citep{Lindegren+2021,Riess+2021}, the PW relation in the \emph{Gaia} bands was calculated as $W_G = (-3.391 \pm 0.052) \log P + (-2.744 \pm 0.045)$ \citep{Ripepi+2022,Ripepi+2022b}. However, the systematic errors of the \emph{Gaia} parallaxes are still quite uncertain for individual stars \citep{Lindegren+2021}, which affect the subsequent measurement accuracies of PW relation. 

By taking the average over a few hundred or thousand stars within an open cluster (OC), the member stars of which have similar properties and are held together by mutual gravitation, one can reduce the potential systematic parallax uncertainties and yield more precise distances compared to using individual parallax measurements \citep{Breuval+2020}. As such, we must ask whether OCs are a useful counterpart to gain independently the precise distances of DCEPs which can provide further calibration of the PW relation? Indeed, DCEPs are probably located in OCs because they are young. As such, starting with the work of \cite{Irwin_1955}, identifying Cepheid variables that are part of OCs has attracted scientific attention of many astronomers and remains an important research topic \citep[][]{Majaess+2008,Turner+2008,Anderson+2013,Medina+2021,Zhou_Chen_2021,Hao_new}. In addition, reliable physical parameters of a DCEP, e.g., age, distance, extinction, and metallicity, can be also obtained by isochrone fitting of its host OC, because the member stars share similar physical properties. These precise distances, metallicities, and extinctions provide external calibration for the PL and PW relations of DCEPs.

Recently, the European Space Agency’s (ESA) \emph{Gaia} mission has provided an third data release (DR3), including astrometry and broadband photometry for a total of 1.8 billion objects \citep{Gaiadr3+2022}. Combined with a large number of DCEPs \citep{Pietrukowicz+2021} and OCs~\citep{Cantat-Gaudin+2018,Cantat-Gaudin+2019,Cantat-Gaudin+2020,Castro-Ginard+2018,Castro-Ginard+2019,Castro-Ginard+2020,Castro-Ginard+2022,Hao+2020,Hao+2022}, a census of open cluster classical Cepheids (OC-DCEPs) is compiled in the present work. With precise distances afforded by cluster-averaged parallaxes, our aim is to provide an independent calibration of the PW relation in the \emph{Gaia} bands.

The remainder of this work is organized as follows. In Section~\ref{sec2}, we introduce our OC-DCEP sample. We calibrate the PW relation of the OC-DCEP sample in Section~\ref{sec3}. Further discussions are presented in Section~\ref{sec4}. Finally, we summarize our principal conclusions in Section~\ref{sec5}.

\section{Sample}
\label{sec2}

Mainly based on the long-term Optical Gravitational Lensing Experiment \citep[OGLE,][]{Udalski+2018,Soszynski+2020}, which surveys field variable stars and searches for Cepheids candidates, and inspected candidates for Cepheids from several published Cepheid surveys \citep[e.g.,][]{Pojmanski+2005,Heinze+2018,Jayasinghe+2020}, \cite{Pietrukowicz+2021} compiled a catalog containing 3352 reliable DCEPs. Besides, \cite{Pietrukowicz+2021} also gave the \emph{Gaia} EDR3 identifier of each DCEP by adopting a matching radius of 0$.''$5. The purity of this catalogue exceeds 97\%, and its completeness is of about 88\% at a magnitude $G = 18$~mag. 

Combined with the newest results of hunting for OCs in the directions of DCEPs \citep{Hao_new}, we performed a cross-match between $\sim$ 3300 DCEPs and $\sim$ 2700 OCs \citep[][]{Cantat-Gaudin+2020,Castro-Ginard+2022} in five-dimensional parameter space ($\alpha, \delta, \varpi, \mu_{\alpha}\cos \delta, \mu_{\delta}$). Finally, we complied 51 OC-DCEP candidates (as shown in Table~\ref{tab1}, and see Appendix~\ref{secA1} for details). Among them, 33 DCEPs are identified as members of OCs \citep[][]{Cantat-Gaudin+2020,Castro-Ginard+2022,Hao_new}, and 18 DCEPs are potential cluster members based on the five membership constraints.

The pulsation modes and periods provided by \cite{Pietrukowicz+2021} imply that our OC-DCEP candidates contain 40 fundamental-mode (F-mode) pulsators, 10 first-overtone (1O mode) pulsators, and one multimode pulsator. To establish a PW relation including the 1O mode OC-DCEP candidates, we adopted $P_{\rm F} = P_{\rm 1O}/(0.716 - 0.027 \log P_{\rm 1O})$ \citep{Feast_Catchpole_1997} to fundamentalise their period. Even though converting 1O mode DCEPs into F-mode DCEPs may introduce a small uncertainty on the derived periods, including the nine first overtones of the sample with their modified periods may improve the precision of the fit.

By integrating the light curve in intensity and then transforming back into magnitude, the corrected average magnitude of a DCEP is beneficial for establishing an accurate PW relation. The vast intensity-averaged apparent magnitudes and errors of the three bands ($G$, $G_{BP}$, and $G_{RP}$) in Table~\ref{tab2} were released in {\texttt gaiadr3.vari\_cepheid} catalogue \footnote{\url{http://archives.esac.esa.int/gaia/}}. Note that DCEP U Sgr is not included in the {\texttt gaiadr3.vari\_cepheid} catalog. We discarded this Cepheid when fitting the PW relation to avoid interference between the arithmetic-averaged magnitude and intensity-averaged magnitude.

\section{Result}
\label{sec3}

\subsection{Distance of OC-DCEP candidates}\label{sec3.1}

For each OC-DCEP candidate, we can obtain its distance using three independent methods. First, for each individual star, \emph{Gaia} DR3 provides its parallax and its uncertainty as $\varpi_{\rm DCEP}$ and $\sigma_{\varpi_{\rm DCEP}}$, respectively. Second, by taking the average over a sample of parallax measurements of an OC population, the host OC can also provide a precise distance of any OC-DCEPs therein. To avoid interference of member stars with a low parallax accuracy ($\sigma_{\varpi} / \varpi$), the cluster-averaged parallax $\varpi_{\rm OC}$ is the mean parallax of member stars with parallax accuracies better than 10\%, where $\sigma_{\varpi_{OC}}$ is the standard deviation of the member star parallaxes. Third, employed with empirical isochrone fitting in color--magnitude diagrams, the distance modulus, $DM$, can also be used to estimate the distance of an OC-DCEP. $DM$ is directly obtained from the corresponding OC catalog. The distance modulus uncertainty, $\sigma_{DM}$, typically ranges from 0.1 to 0.2 mag~\citep{Cantat-Gaudin+2020}. We simply set $\sigma_{DM}$ to 0.1 mag when the member stars of the OC numbers more than 100, and 0.2 mag for OCs with fewer members.

\begin{figure}[!ht]
  \centering
  \includegraphics[width=0.32\textwidth]{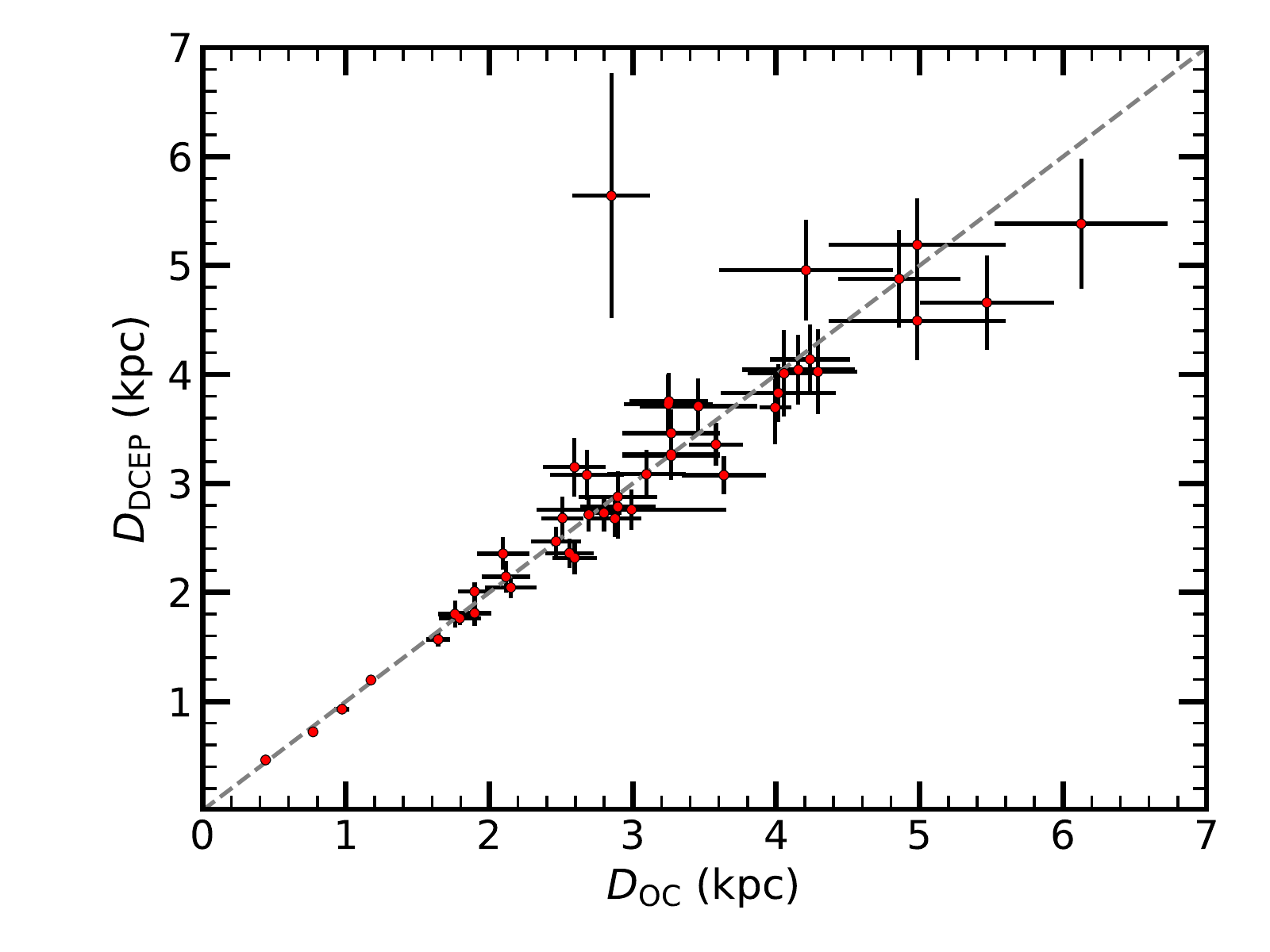}
  \includegraphics[width=0.32\textwidth]{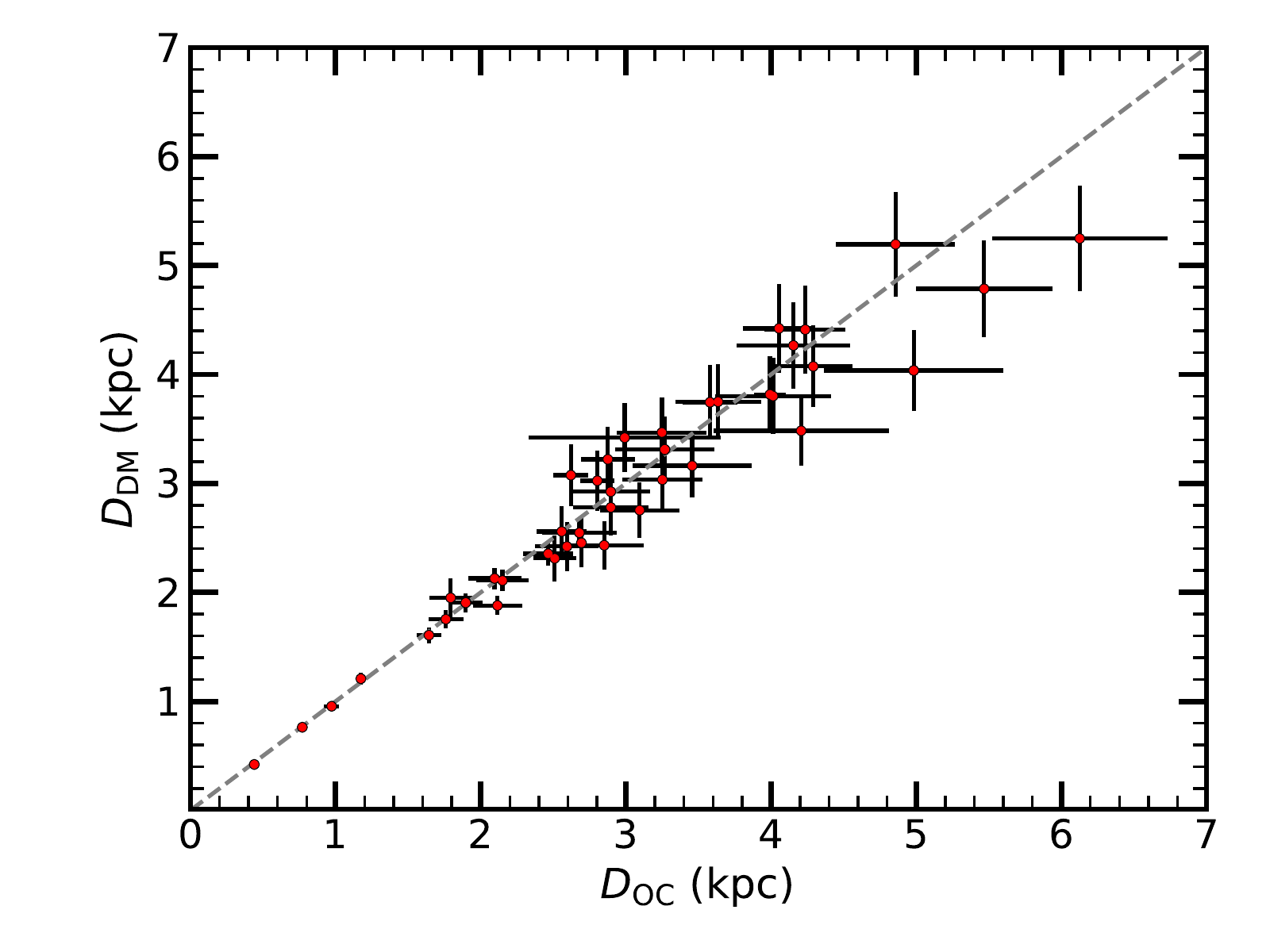}
  \includegraphics[width=0.32\textwidth]{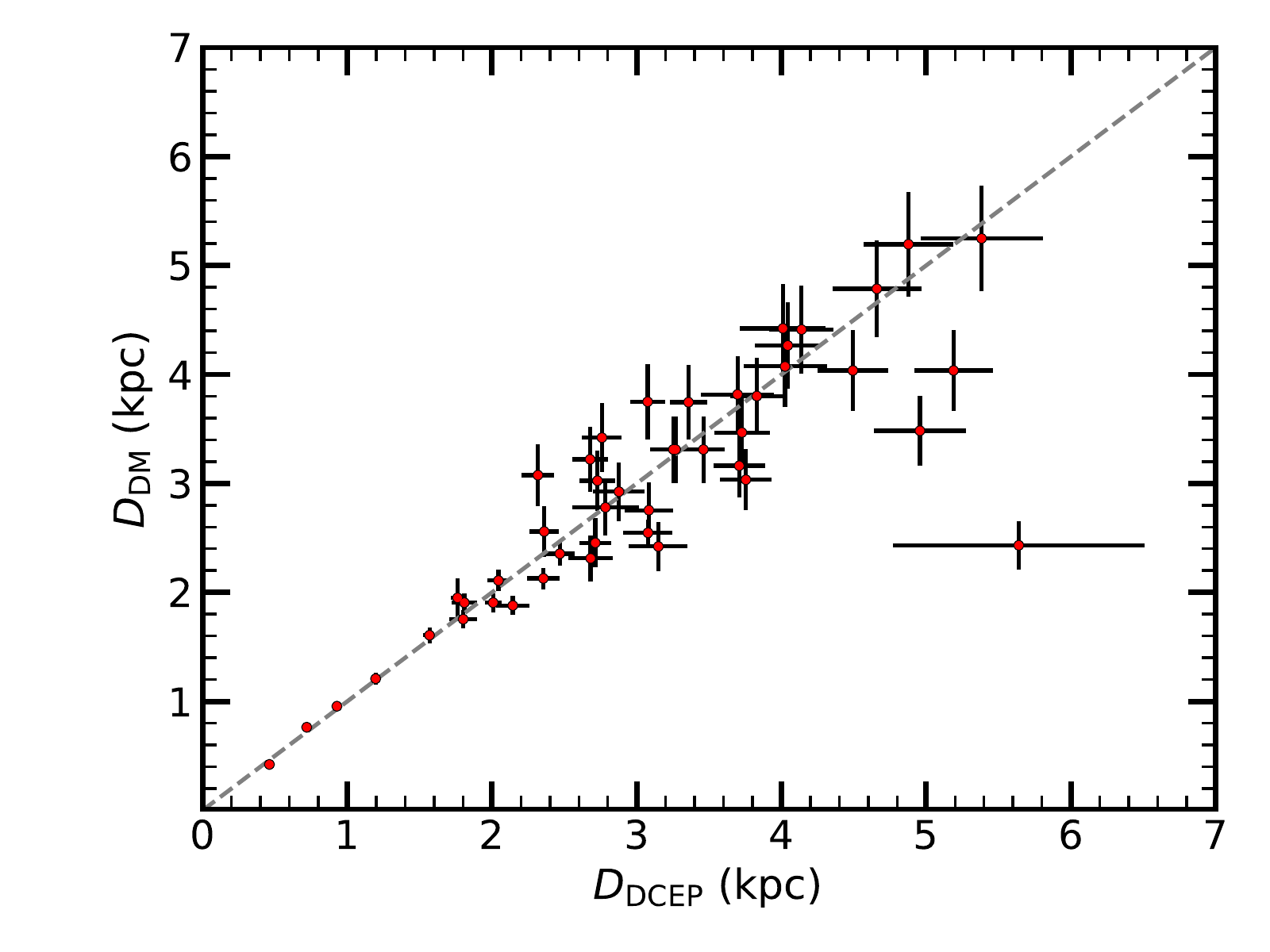}
  \caption{Left panel: a comparison of distances between adopting the cluster-averaged parallax and using the individual DCEP parallaxes. Middle panel: a comparison of distances between the cluster-averaged parallaxes and the distance moduli. Right panel: a comparison between the DCEP parallax distances and the distance modulus distances. The dashed line corresponds to the identity line.}
  \label{fig1}
\end{figure}

The measured parameters for each OC-DCEP candidate are summarized in Table~\ref{tab2}. The three panels in Figure~\ref{fig1} present pairwise comparisons of the distances obtained by the three methods. The distances obtained by the three methods are mostly consistent.

\subsection{Period-Wesenheit relation}

For the three bands of \emph{Gaia} ($G, G_{BP}, G_{RP}$), according to the assumed extinction law and coefficient, the PW relation and reddening-free Wesenheit apparent magnitude, $w_G$, can be described as \citep{Madore_1982,Madore+2017},

\begin{align}
W_G &= a~\log P + b \\
w_G &= m_G - \lambda \times (m_{G_{BP}} - m_{G_{RP}}) \label{eq2}\\
W_{G_{obs}} &=  w_G + 5 \times \log(\varpi / mas) - 10 \label{eq3}
\end{align}

\noindent where, $W_G$ is Wesenheit absolute magnitude in \emph{Gaia} $G$ band, $W_{G_{obs}}$ is observed Wesenheit absolute magnitude obtained by Wesenheit apparent magnitude and trigonometrical parallax ($\varpi$), $P$ is period, and $\lambda = A(G)/E(G_{BP}-G_{RP})$.

\begin{figure}[!ht]
    \centering
    \subfigure[]{\includegraphics[width=0.32\textwidth]{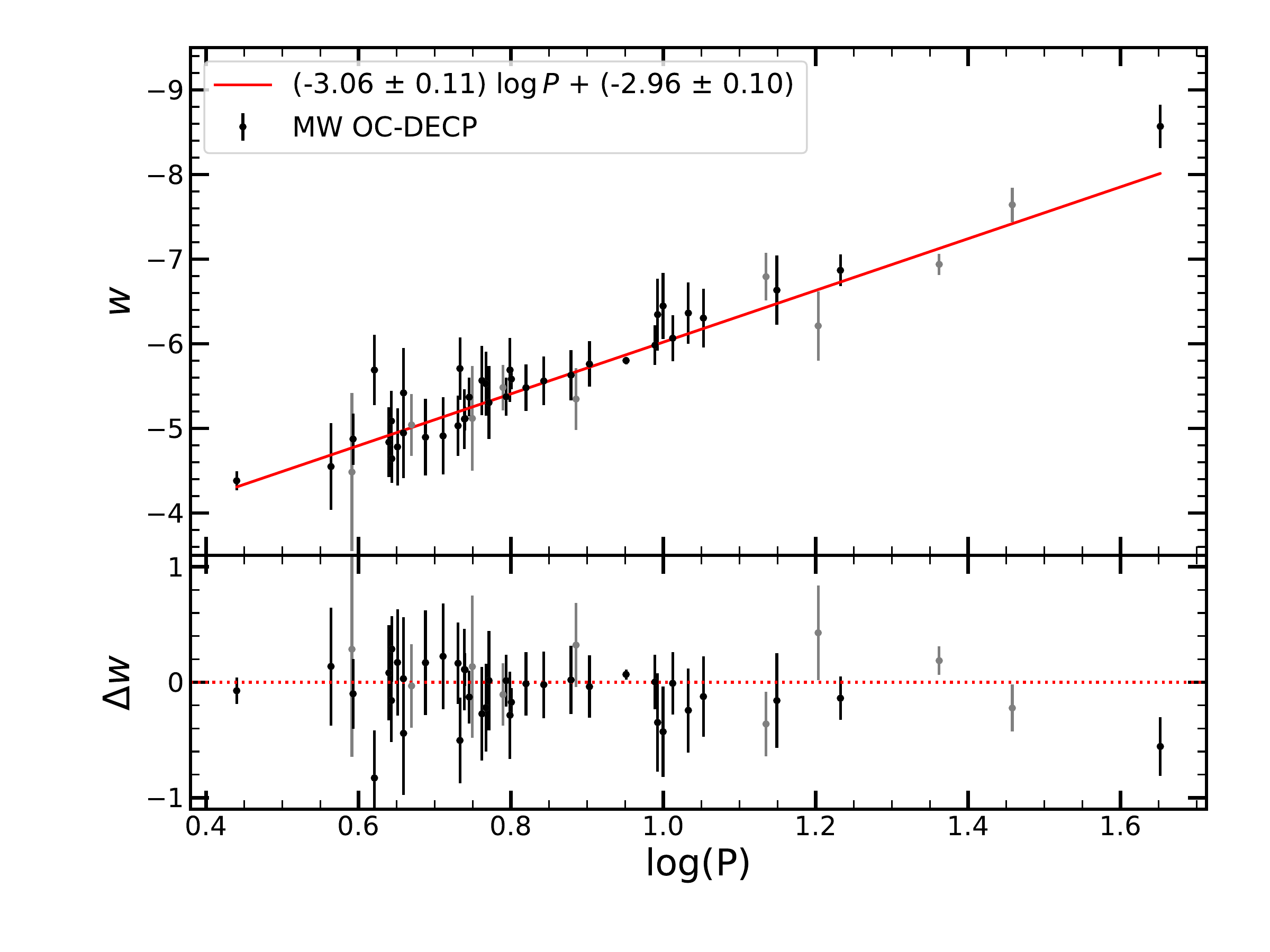}}
    \subfigure[]{\includegraphics[width=0.32\textwidth]{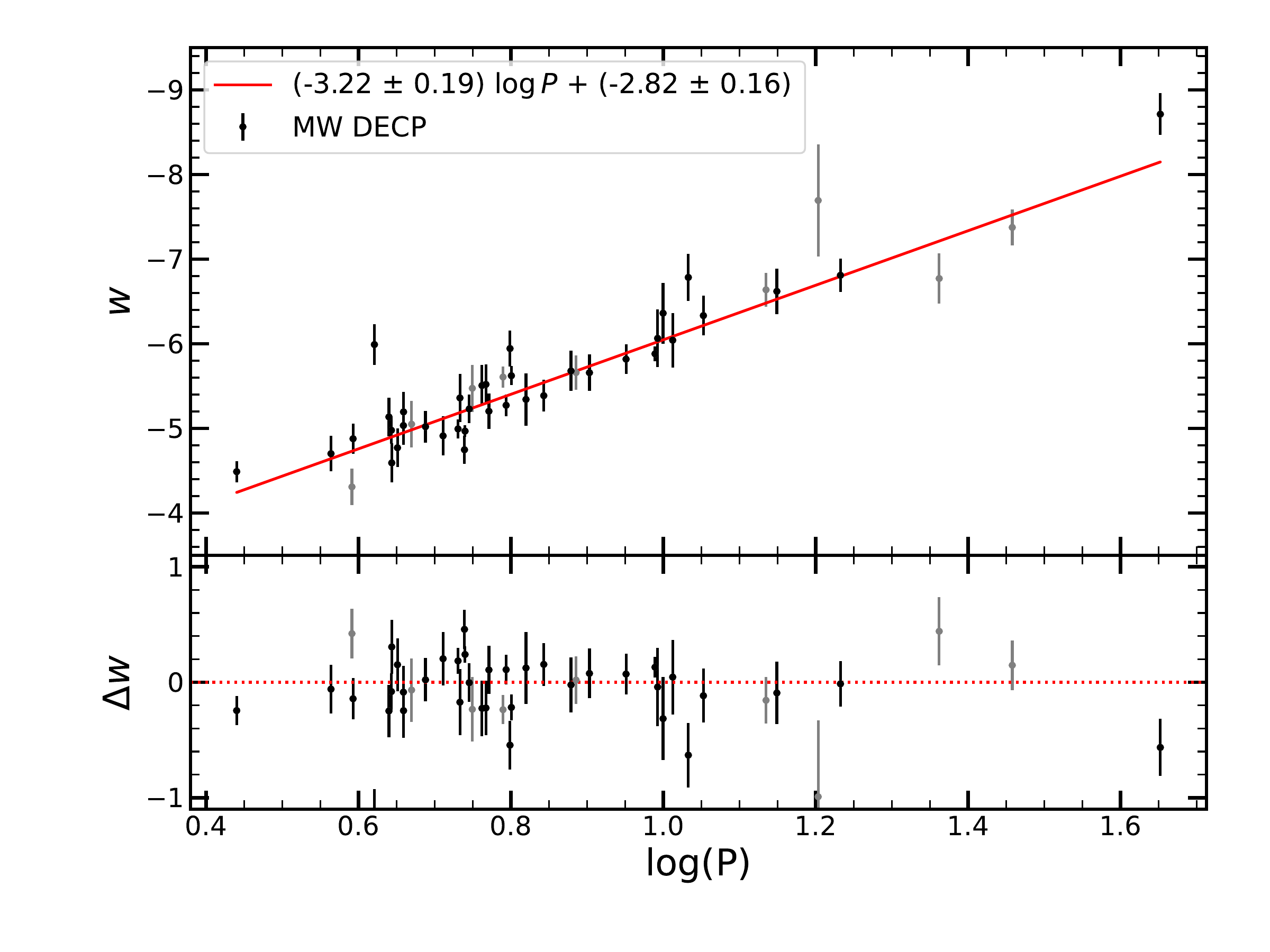}}
    \subfigure[]{\includegraphics[width=0.32\textwidth]{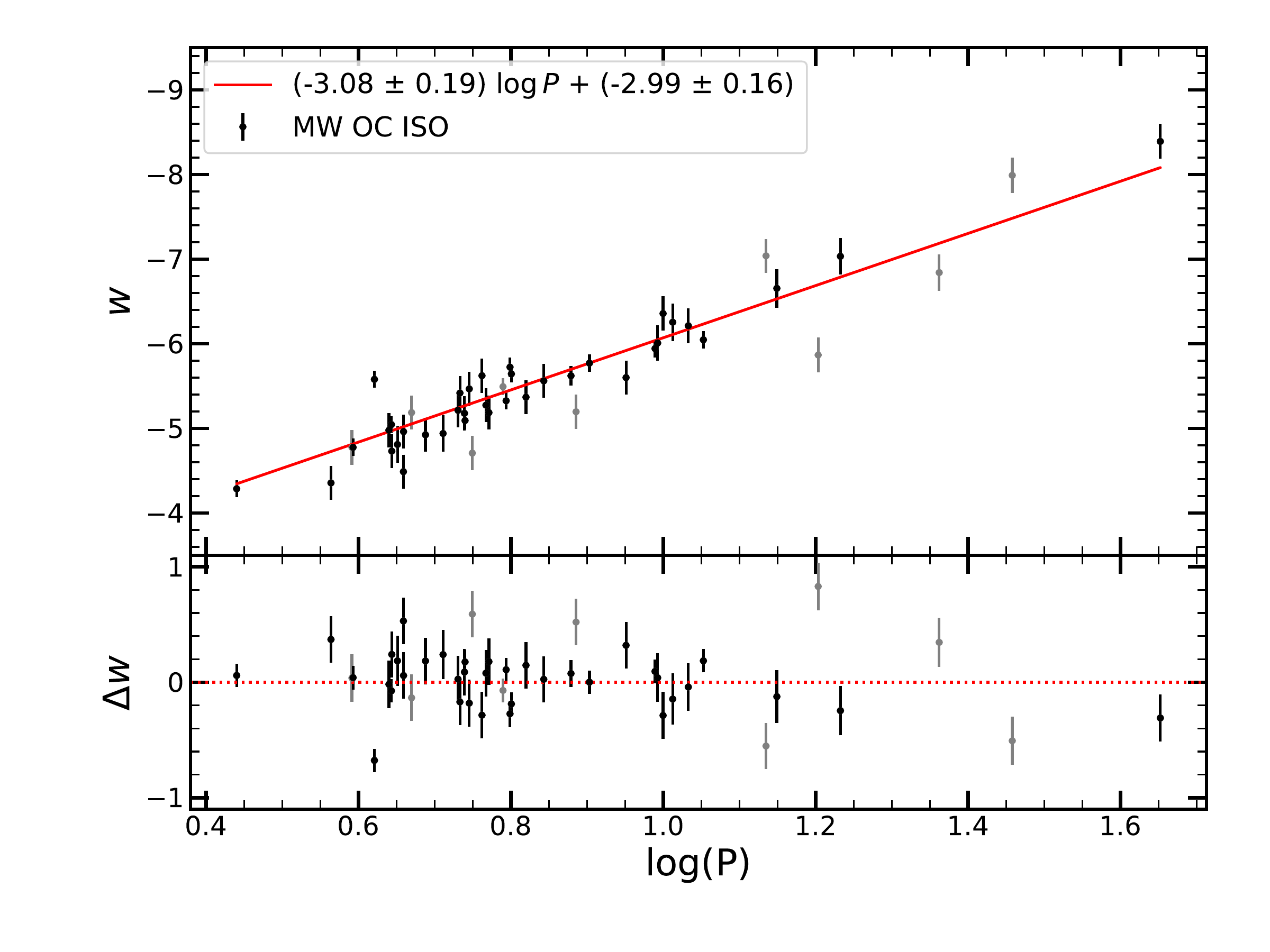}}
    \subfigure[]{\includegraphics[width=0.32\textwidth]{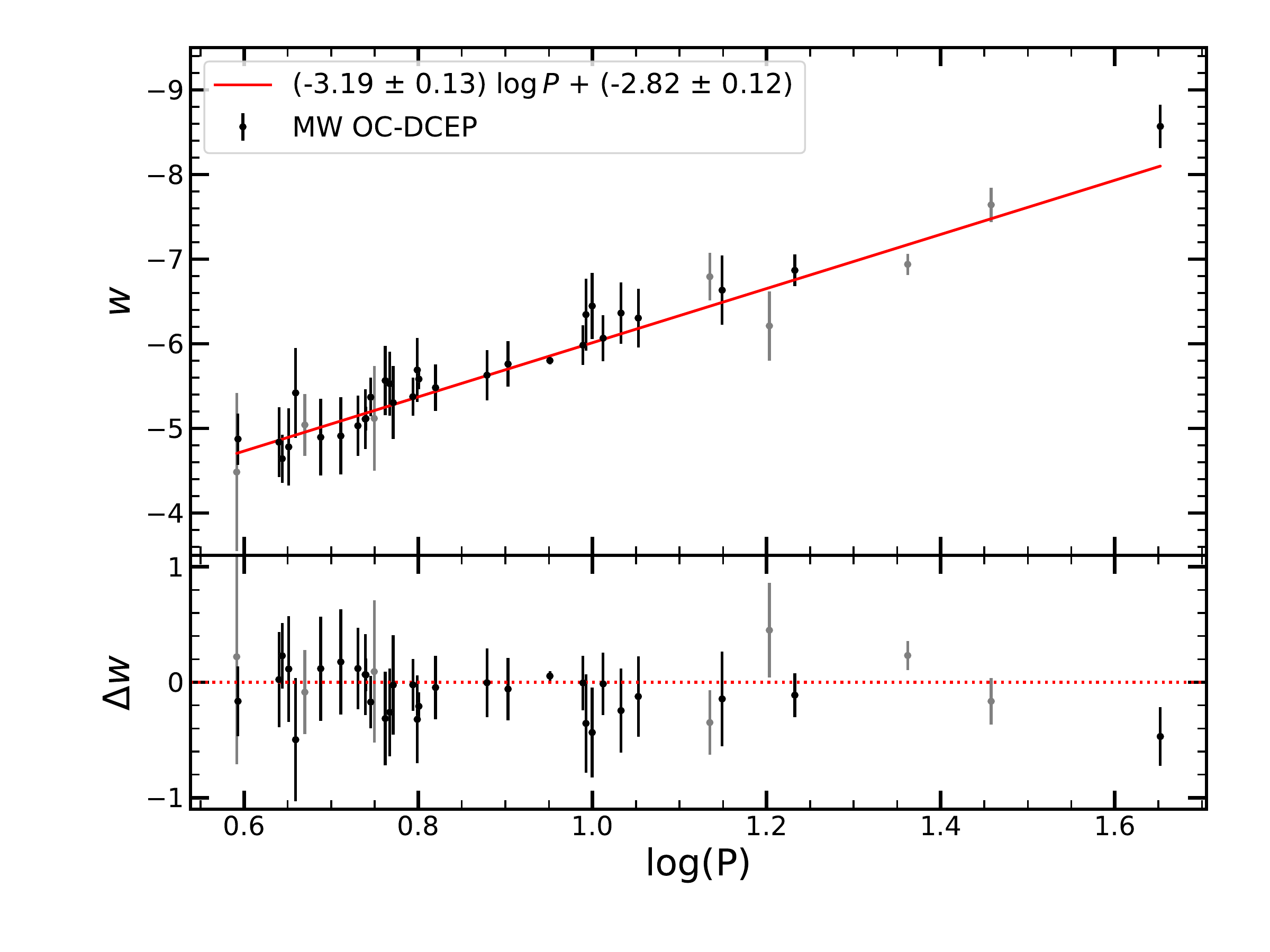}}
    \subfigure[]{\includegraphics[width=0.32\textwidth]{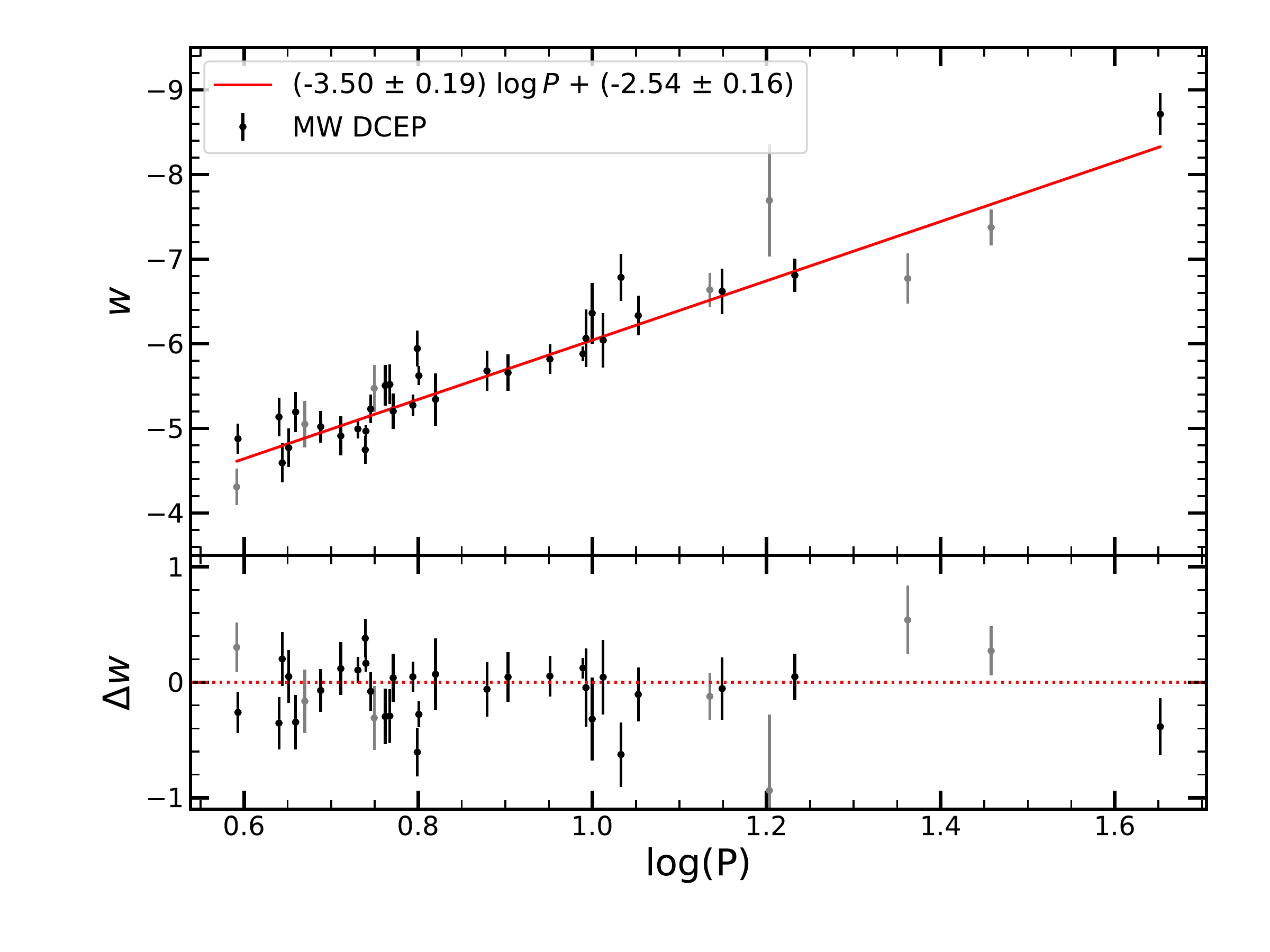}}
    \subfigure[]{\includegraphics[width=0.32\textwidth]{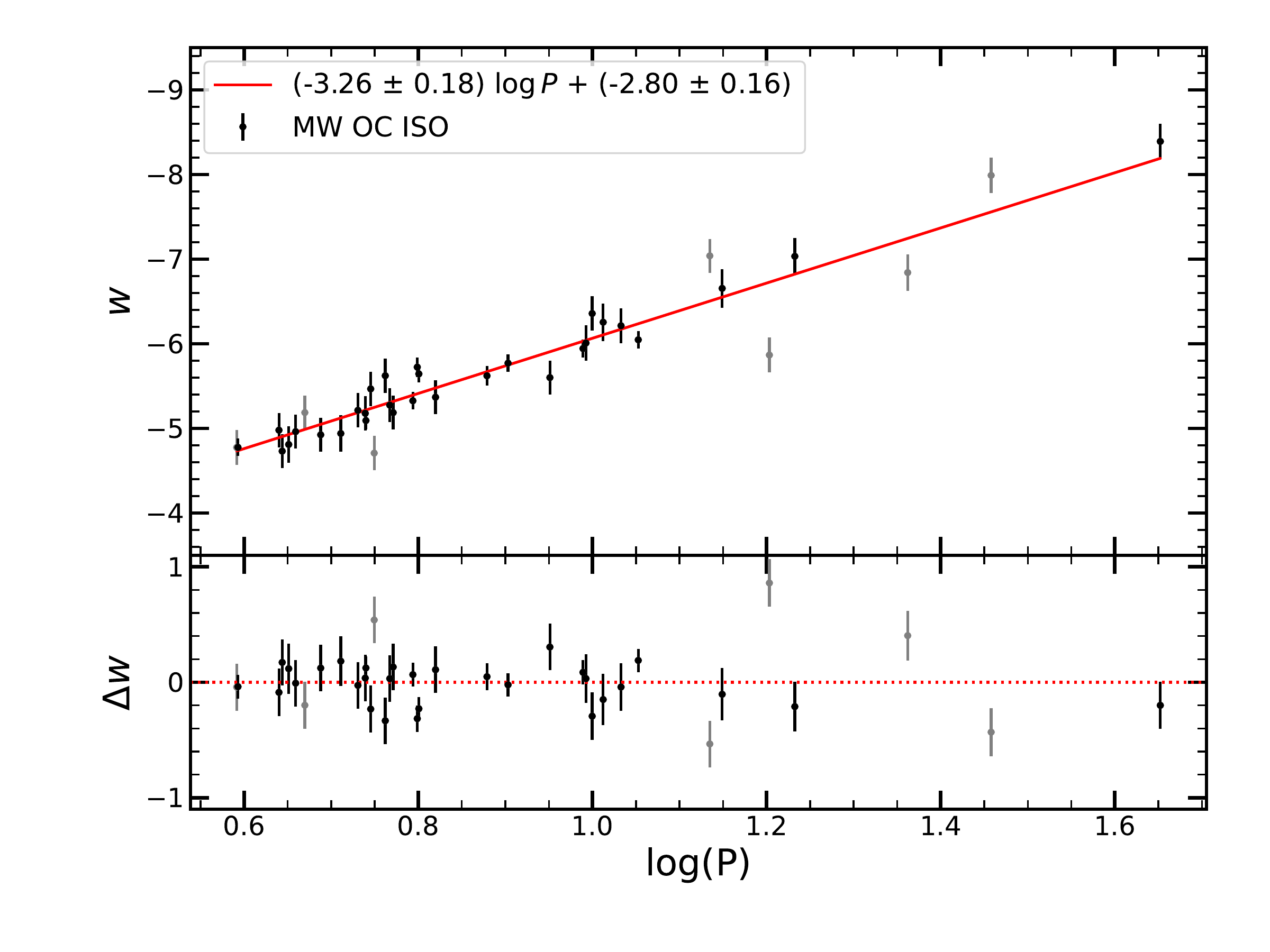}}
    \caption{PW relation of the OC-DCEP candidates in the $G$-bands. From left to right, the top three panels are: PW relations obtained by fitting the F+1O mode OC-DCEP candidates, where the distances are adopted from the cluster-averaged parallaxes (a), DCEP parallaxes (b), and distance moduli (c), respectively. The bottom three panels are similar to the top panels, but the PW relations were obtained by fitting just the F-mode OC-DCEP candidates. The gray dots present the extended OC-DCEP candidate sample with several dimensions outside 3$\sigma$, but less than 4$\sigma$, in the five-dimensional parameter space.}
    \label{fig2}
  \end{figure}

In this section, we simply adopt the empirical $\lambda = 1.9$ \citep{Ripepi+2019}. Only OC-DCEP candidates who have three independent distances were used. The total number of F+1O mode and F-mode OC-DCEP candidates adopted in fitting the PW relation are 46 and 37, respectively. The results of the PW relation are shown in Table~\ref{tab4} and Figure~\ref{fig2}, obtained by standard least-squares fitting.

For the OCs, the parallax averaged over a large number of stars at similar distances can reduce the measurement errors caused by individual objects, to some extent, and obtain a more precise distance. The fitted PW relation is $W_G = (-3.06 \pm 0.11) \log P + (-2.96 \pm 0.10)$ by using 46 OC-DCEP candidates with a cluster-averaged parallax. The results of the PW relation by adopting the cluster-averaged parallaxes show the largest correlation coefficient ($R$) and smallest dispersion ($\sigma_{fit}$), indicating that it is more appropriate to use the cluster-averaged parallaxes to calibrate PW relation.

The slope ($a$) and zero-point ($b$) of the PW relation obtained by using the parallaxes of the F+1O mode DCEPs are consistent with the results of fitting the PW relation by directly using the DCEPs' individual parallax results in the \emph{Gaia} bands~\citep[$a$ = -3.289 $\pm$ 0.039, $b$ = -2.739 $\pm$ 0.013; see][]{Ripepi+2022}. Meanwhile, the slope of the PW relation calculated by the cluster-averaged parallaxes is shallower than the result by using the individual DCEP parallaxes, and the value of zero-point is also smaller. Different slopes and zero-points of the PW relation obtained with \emph{Gaia} DR2 parallaxes of companions and OCs and with direct parallaxes of DCEPs were calibrated in the $V$, $J$, $H$, $K_S$, and Wesenheit $W_H$ bands by \citep{Breuval+2020}. Adopting the same Wesenheit apparent magnitudes, $w_G$, the differences in the PW relation are primarily due to the distances of the OC-DCEP candidates being obtained by different methods (as shown in Figure~\ref{figA1} and Section~\ref{sec3.1}). Indeed, due to the uncertain systematic errors of the parallaxes \citep{Lindegren+2021} and the underestimated formal uncertainties \citep{Fabricius+2021}, the measured parallax accuracies of individual DCEP may be highly impacted. The effects of the systematic errors on the DCEP parallaxes should be reduced in following \emph{Gaia} data releases thanks to the large number of measurements that will be contained therein. For a cluster, a large number of member stars can reduce the influence of systematic errors and yield more precise distances.

\begin{deluxetable}{lcccc}[!ht]
  \tablecolumns{5}
  \tabletypesize{\normalsize}
  \setlength\tabcolsep{20pt}
  \tablecaption{PW Relation of OC-DCEPs with Different Distances\label{tab4}}
  \tablehead{
  \colhead{} & \colhead{$a$}  & \colhead{$b$} & \colhead{$\sigma_{fit}$} & \colhead{$R$} 
  }
  \startdata
  MW OC-DCEP   & -3.06 $\pm$ 0.11 & -2.96 $\pm$ 0.10 & 0.25 & 0.95 \\
  MW DCEP      & -3.22 $\pm$ 0.19 & -2.82 $\pm$ 0.16 & 0.33 & 0.93 \\
  MW OC ISO    & -3.08 $\pm$ 0.19 & -2.99 $\pm$ 0.16 & 0.29 & 0.94 \\
  \hline
  MW OC-DCEP F & -3.19 $\pm$ 0.13 & -2.82 $\pm$ 0.12 & 0.21 & 0.97 \\
  MW DCEP F    & -3.50 $\pm$ 0.19 & -2.54 $\pm$ 0.17 & 0.29 & 0.93 \\
  MW OC ISO F  & -3.26 $\pm$ 0.18 & -2.80 $\pm$ 0.16 & 0.26 & 0.95 \\
  \enddata
  \tablecomments{The PW relation is fitted to the OC-DCEP sample using the standard least-square method, with a fixed $\lambda$ = 1.9. Among them, the MW OC-DCEPs show the results of the F+1O mode OC-DCEP candidates by using $\varpi_{\rm OC}$ as its distance; MW DCEP is the PW relation obtained by fitting the F+1O mode OC-DCEP candidates with the distance adopted with $\varpi_{\rm DCEP}$; and MW OC ISO presents the PW relation when using the $DM$. The PW relations shown in the last three rows are similar with MW DCEP, MW OC-DCEP, and MW OC ISO, but are only for the F-mode OC-DCEP candidates. $\sigma_{fit}$ is the standard deviation of the differences between the observed absolute Wesenheit magnitudes and the Wesenheit magnitudes calculated from the coefficients $a$ and $b$ reported in the table. $R$ is the correlation coefficient.}
  \end{deluxetable}
  
Isochrone fitting procedures can also independently measure the photometric distance of a cluster, i.e., obtained by the $DM$. The PW relation of the OC-DCEP candidates by using the $DM$ is consistent with the result of adopting $\varpi_{\rm OC}$. The external calibration of the distance moduli also points to a shallower slope and smaller zero-point value compared with the calibration of individual DCEP distances. In many cases, the photometric distances have strong model- and parameter-dependence, and may be affected by field contamination and degeneracy between distances, reddening, and metallicity. The fitted results of $R$ and $\sigma_{fit}$ are not as good as those obtained when using the cluster-averaged parallax.
By using the cluster-averaged parallaxes or distance moduli, the slope is very consistent between the F-mode and F+1O mode DCEPs, for which the discrepancies are $\sim$ 1$\sigma$. Including the 1O mode of the sample with their fundametalised periods instead of rejecting them introduces only a small change in the PW relation and improves the precision of the fit. Meanwhile, when adopting the DCEP parallaxes, the slope of the PW relation obtained by only the F-mode OC-DCEP candidates is steeper than that obtained from the F+1O mode candidates. Previous studies based on a larger sample of DCEPs have shown that the PW relationships obtained by F-mode and F+1O mode DCEPs have good consistency \citep{Ripepi+2019,Ripepi+2022}. This also implies that using the cluster-averaged parallax is more suitable for calibrating the PW relation of OC-DCEP candidates. The deviation of the slope between the fit obtained by using the averaged parallax of the host OC and directly measured DCEP parallaxes is more than 1$\sigma$ for both the F-mode and F+1O mode OC-DCEP candidates. Given the limited size of the OC-DCP sample used in this work, more objects and precise distances are needed for to calibrate further the PW relation using future \emph{Gaia} data releases.

\section{Discussion}
\label{sec4}

\subsection{Selection effects}

In order to avoid selection effects of the cluster-averaged parallax, $\varpi_{\rm OC}$, caused by selecting high-precision (10\%) member stars, we also tested (1) selecting member stars with a parallax accuracy better than 20\% and (2) placing no limitation on the parallax accuracy. 
Whether we limited the parallax accuracy of member stars or not, the cluster-averaged parallaxes were always consistent. The slope and zero-point of the PW relation fitted by using the limited (20\%) or unlimited members are also consistent with the result afforded by the 10\% parallax accuracy member stars, but a more precise PW relation with a lower uncertainty was obtained by using just the high-precision member stars.

\begin{deluxetable}{lccc}[!ht]
  \tablecolumns{4}
  \tabletypesize{\normalsize}
  \setlength\tabcolsep{20pt}
  \tablecaption{PW Relation of Different Samples by Adopting the OC Parallax Distances\label{tab5}}
  \tablehead{
  \colhead{} & \colhead{N} & \colhead{$a$}  & \colhead{$b$}
  }
  \startdata
  all & 46  & -3.06 $\pm$ 0.11 & -2.96 $\pm$ 0.10 \\
  $<$ 3 $\sigma$ & 37 & -3.12 $\pm$ 0.13 & -2.91 $\pm$ 0.12 \\
  members & 30 & -3.13 $\pm$ 0.14 & -2.90 $\pm$ 0.12 \\
  \enddata
  \tablecomments{With a fixed $\lambda$ = 1.9 and standard least-square fitting, the PW relations were modified by using different samples. The second row ($<$3$\sigma$) is a sample containing DCEPs whose five-dimensional parameters are consistent with the five-dimensional parameters of the associated OC within 3$\sigma$. The third row (members) shows the result for the sample where the DCEPs are identified as member stars of the host OC.}
\end{deluxetable}

As described in Section~\ref{sec3}, the PW relation result fitted by directly using the individual parallaxes of the DCEPs is different from that obtained by the cluster-averaged parallaxes. To describe the effect of using a sample of identified membership and that of potential membership of OC-DCEP candidates qualitatively, we compared the fitted PW relations of samples where the five-dimensional parameter space of the DCEPs and OCs were consistent to within 3$\sigma$, and samples where the DCEPs were identified members of the host OC. The results of three samples, including all OC-DCEP candidates, are very consistent (as shown in Table~\ref{tab5}). When using DCEPs that are member stars of OCs, the derived PW relation also shows a shallower slope and smaller zero-point value than the known results. Moreover, there is no obvious deviation between the grey points (extended OC-DCEP candidate sample) and fitted PW relation in Figure~\ref{fig2}. Therefore, there is little effect when using a sample of identified membership and sample of potential membership of OC-DCEP candidates. Including potential OC-DCEP candidates, and by increasing the total amount of OC-DCEPs in the sample by 50\% relative to the identified membership sample, can help establish a more precise PW relation.

\subsection{Parallax Zero-point Correction} \label{sec4_2}

\begin{figure}[!ht]
  \centering
  \subfigure[]{\includegraphics[width=0.48\textwidth]{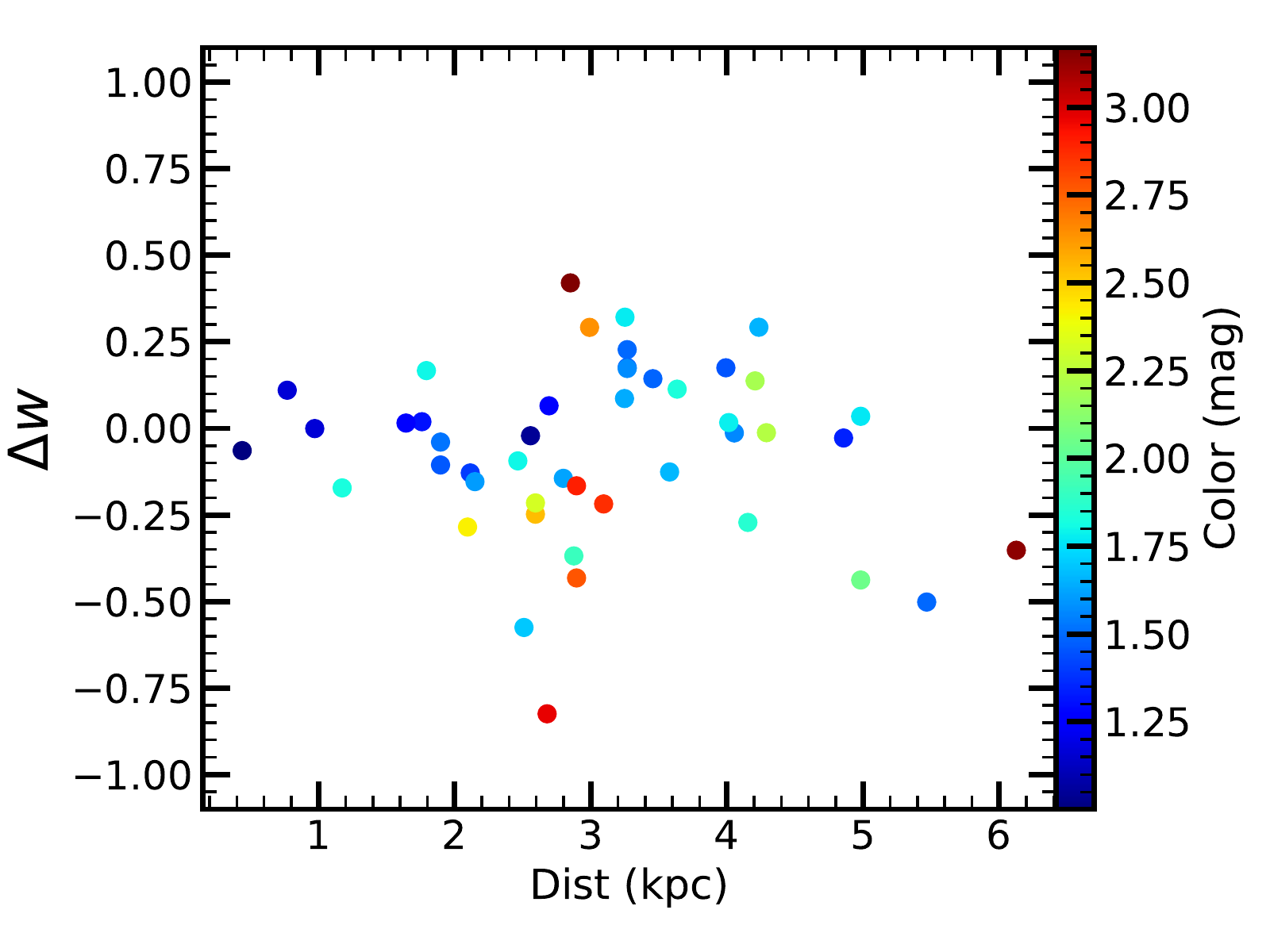}}
  \subfigure[]{\includegraphics[width=0.48\textwidth]{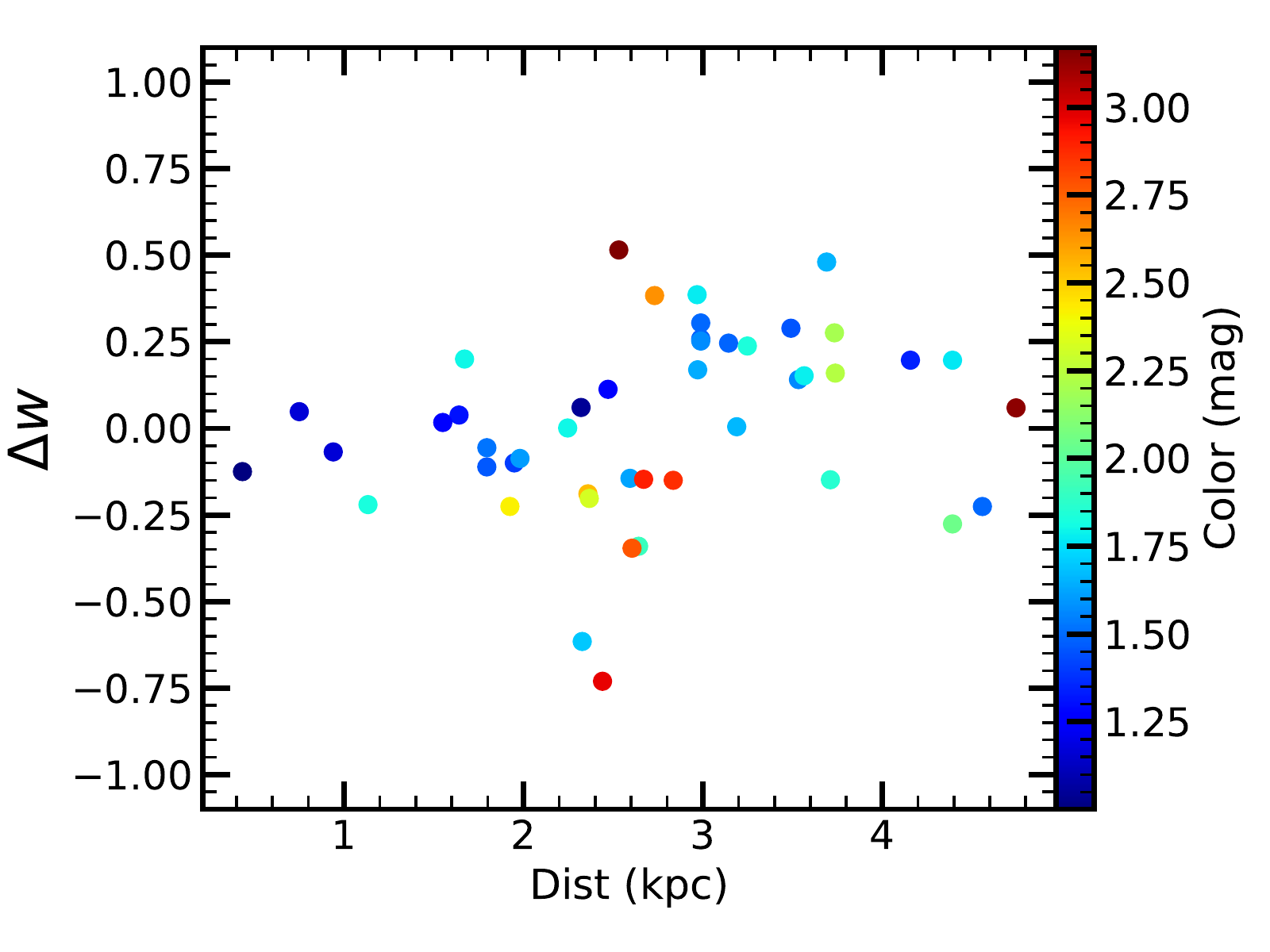}}
  \caption{Deviation between the observed $W_{G_{obs}}$ and the $W_{G}$ obtained from the PW relation (i.e., $\Delta W = W_{G_{obs}}-W_{G}$) and distance, where the different colors denote the colors of the OC-DCEP candidates. The samples with and without using parallax zero-point correction are shown in panels (a) and (b), respectively.}
  \label{fig3}
\end{figure}

\begin{figure}[!ht]
  \centering
  \includegraphics[width=0.48\textwidth]{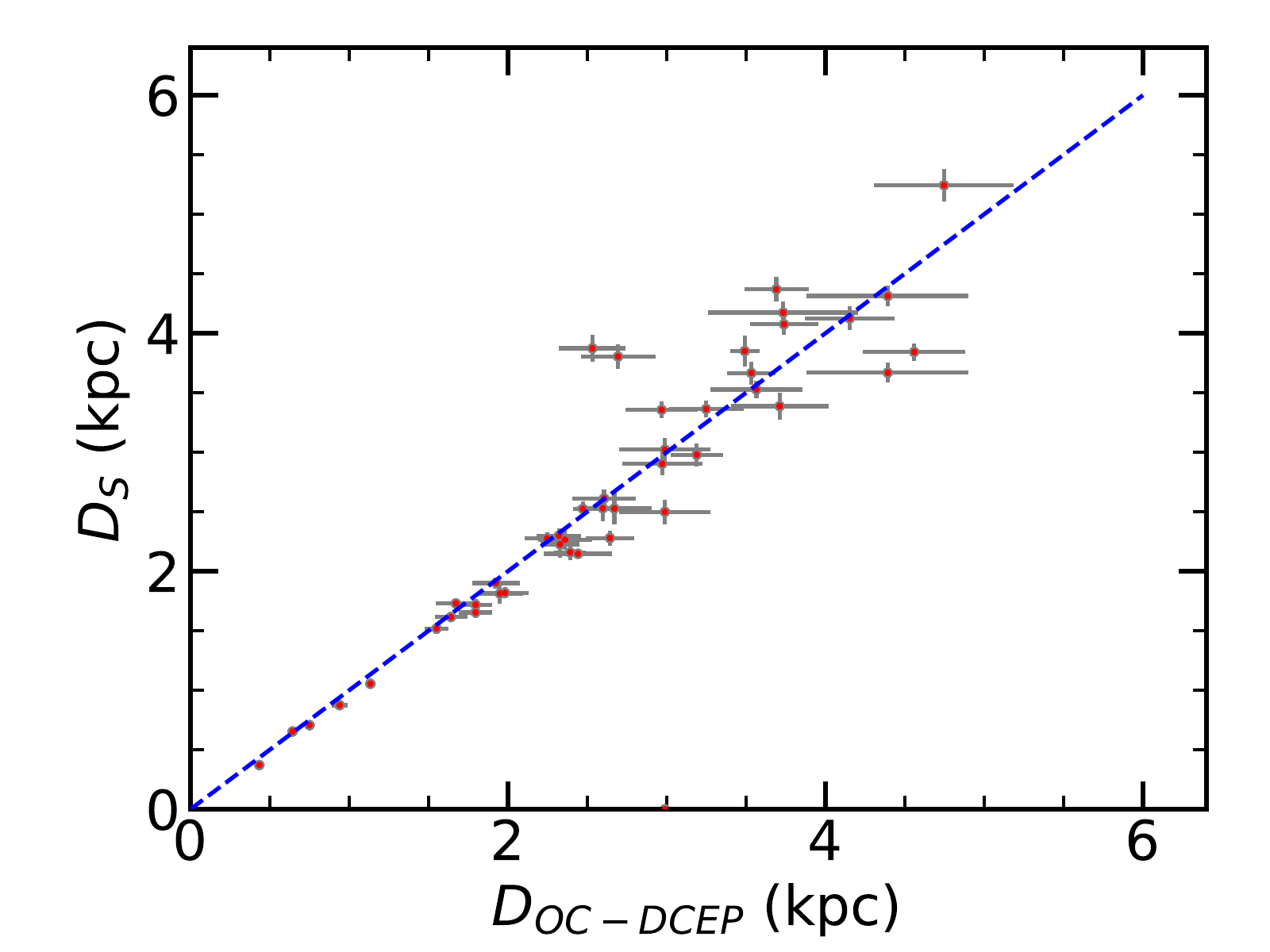}
  \caption{Comparison of OC-DCEPs distance between provided by~\cite{Skowron+2019}, $D_S$, and obtained by cluster-averaged parallax after considering parallax zero-point correction. The dashed line corresponds to the identity line.}
  \label{fig_compare}
\end{figure}

Figure~\ref{fig3} shows the deviations between the observed $\mathbf{ W_{G_{obs}}}$ and the $\mathbf{ W_{G}}$ obtained from the PW relation (i.e.,  $\mathbf{ \Delta W = W_{G_{obs}}-W_{G}}$) at different distances, where the different colors denote the color, $m_{G_{BP}} - m_{G_{RP}}$, of the OC-DCEP candidates. As shown in panel (a) in Figure~\ref{fig3}, the $W_{G_{obs}}$ values tend to be smaller than $W_{G}$ at large distances. Based on the Equation~\ref{eq3}, a smaller $W_{G_{obs}}$ suggests that the parallax may be underestimated. 
Considering that the averaged parallaxes of the OCs may also have a negative systematic offset \citep{Dias+2021}, we used a parallax zero-point correction \citep{Lindegren+2021} for each member star and then averaged their parallaxes, as done for the OC parallaxes.
Figure~\ref{fig3}(b) shows the distributions of distance and $\Delta W$ with a parallax zero-point correction. There is no obvious trend where $W_{G_{obs}}$ is smaller than $W_{G}$ at large distances. 
This suggests that using a zero-point correction \citep{Lindegren+2021} for member stars can contribute to reducing the negative systematic offset of the OC parallaxes corroborating the latest result reported by~\cite{Riess+2022}. After adopting parallax zero-point correction for cluster Cepheids, \cite{Riess+2022} found no evidence of residual parallax offset.
Meanwhile, as shown in Figure~\ref{fig_compare}, the distances to Cepheids derived from the zero-point corrected OC parallaxes are well agreement with those obtained from the classical PL relation provided by~\cite{Skowron+2019}. Hence, we propose that the PW relation derived from the Cepheid distances found by the zero-point corrected OC parallax may be more reliable.
The results of the PW relation by using the averaged parallax with or without a parallax zero-point correction are shown in Table~\ref{tab6}. Remarkably, when the parallax zero-point correction was adopted, the slope of the PW relation became shallower than without using parallax zero-point correction.

\begin{deluxetable}{lccc}[!ht]
\tablecolumns{4}
\tabletypesize{\normalsize}
\setlength\tabcolsep{20pt}
\tablecaption{PW Relation of the 46 OC-DCEP Candidates Adopted from the Averaged Parallax without or with using Zero-point Correction\label{tab6}}
\tablehead{
\colhead{} & \colhead{$a$} & \colhead{$\lambda$}  & \colhead{$b$}
}
\startdata
 & -3.06 $\pm$ 0.11 & 1.9 & -2.96 $\pm$ 0.10 \\
MW OC-DCEP & -2.96 $\pm$ 0.09 & 1.71 $\pm$ 0.05 & -2.76 $\pm$ 0.11 \\
 & -3.32  & 1.9 & -2.72 $\pm$ 0.03 \\
\hline
 & -2.94 $\pm$ 0.12 & 1.9 & -2.93 $\pm$ 0.11 \\
MW OC-DCEP with zpt & -2.89 $\pm$ 0.12 & 1.81 $\pm$ 0.07 & -2.82 $\pm$ 0.14 \\
 & -3.32  & 1.9 & -2.59 $\pm$ 0.04 \\
\enddata
\tablecomments{The top three rows (MW OC-DCEP) exhibit the fitted results of the PW relation by using the averaged parallaxes without a parallax zero-point correction, and the bottom three rows (MW OC-DCEP with zpt) show the results with a parallax zero-point correction.}
\end{deluxetable}

Similar to \cite{Zhou_Chen_2021}, we tried to fix the slope in the PW relation \citep[$a = -3.32$,][]{Ripepi+2019} and only calibrate the zero-point. The results are also shown in Table~\ref{tab6}. When the parallax zero-point correction is not considered, the zero-point of the PW relation obtained by fitting with a fixed slope is -2.72 $\pm$ 0.03, which is consistent with the result of \cite{Zhou_Chen_2021} but the uncertainty is smaller. As mentioned above, when discussing more distant OC-DCEPs, there is a systematic offset between $W_{obs}$ and $W_{model}$ when a parallax zero-point correction is not used. Applying a parallax zero-point correction will reduce this systematic error. Meanwhile, after considering the parallax zero-point correction, the zero-point of the PW relation is -2.59 $\pm$ 0.04. This exhibits a discrepancy of $\sim$0.13~mag for the samples with and without a parallax zero-point correction.

Considering the average extinction law for diffuse and low-density regions in the Galaxy of $R_V$ = 3.1~\citep{Cardelli+1989}, which is commonly used to correct observations for dust extinction, the value of $\lambda$ is of the order of 2 over a wide range of effective temperatures, including those typically spanned by DCEPs \citep{Andrae+2018}. After a few experiments by testing the Wesenheit magnitude of DCEPs in the LMC, \cite{Ripepi+2019} realized that if the $\lambda$ value is slightly reduced to 1.90, the least square fitting of the DCEPs gives a tighter PW relation. It is worth noting that the differences of $R_V$ between the Milky Way and Magellanic Clouds \citep{Gordon+2003} may cause different values of $\lambda$. Besides, the optical extinction law in the Milky Way exhibits significant diversity \citep{Wang+2017,Fitzpatrick+2019} and has uncertainties \citep[$R_V$ = 3.1 $\pm$ 0.1,][]{Wang_Chen_2019}. Thus, we refit the PW relation with an unfixed $\lambda$. As shown in Table~\ref{tab6}, the best-fitted $\lambda$ obtained without a parallax zero-point correction is $1.71 \pm 0.05$, which deviates from 1.9 by $\sim$3.8 times its uncertainty. By applying a parallax zero-point correction, the best-fitting $\lambda$ also has a deviation of $\sim$1.4$\sigma$ from 1.9.

\section{Conclusions}
\label{sec5}

We compiled 51 OC-DCEP candidates including 33 candidates whose DCEPs are members of a host OC. By adopting different distance indicators, such as the parallaxes of each DCEP, the cluster-averaged parallaxes, and the distance moduli, we gauged the PW relations of the OC-DCEP candidates.
The fitted PW relation obtained by using the cluster-averaged parallaxes is $W_G = (-3.06 \pm 0.11) \log P + (-2.96 \pm 0.10)$, and the obtained PW relation has a tighter correlation and smaller dispersion. This PW relation is different from that obtained directly by using the individual DCEP parallaxes, with a shallower slope and a smaller value of zero-point. The more distant OC-DCEP sample shows a systematic offset between the observed Wesenheit magnitude and fitted Wesenheit magnitude. Applying a parallax zero-point correction \citep{Lindegren+2021} to OC member stars reduces this systematic offset. The PW relation is calibrated as $W_G = (-2.94 \pm 0.12) \log P + (-2.93 \pm 0.11)$. When using a fixed slope ($a$ = -3.32) and fixed $\lambda$ ($\lambda$ = 1.9), the zero-point of the PW relation in \emph{Gaia} bands yields a discrepancy of $\sim$ 0.13 mag between samples with and without a parallax zero-point correction. Identifying more OC Cepheid samples and/or obtaining more precise distance, with a much lower level of systematics in the future, will allow a more precise calibration of the PW relation.

\acknowledgments
We thank the anonymous referee for the helpful comments and suggestions. 
This work was funded by the NSFC Grands 11933011, and 11873019, the Natural
Science Foundation of Jiangsu Province (Grants No. BK20210999) and the
Key Laboratory for Radio Astronomy. L.Y.J.
thanks the support of the Entrepreneurship and Innovation
Program of Jiangsu Province. This work has made use
of data from the European Space Agency(ESA) mission Gaia
(\url{https://www.cosmos.esa.int/gaia}), processed by the Gaia Data
Processing and Analysis Consortium (DPAC, \url{https://www.cosmos.esa.int/web/gaia/dpac/consortium}). Funding for the DPAC has
been provided by national institutions, in particular the institutions
participating in the Gaia Multilateral Agreement.

\begin{appendix}

\section{OC-DCEP sample}\label{secA1}
\setcounter{table}{0}
\setcounter{figure}{0}
\renewcommand{\thetable}{A\arabic{table}}
\renewcommand{\thefigure}{A\arabic{figure}}

Since \cite{Irwin_1955} found two DCEPs associated with OCs, a lot of work has been devoted to searching for Cepheids and cluster combos in terms of their spatial distributions, motions, and other dimensions \citep{Majaess+2008,Anderson+2013,Turner+2008,Medina+2021,Zhou_Chen_2021}. \cite{Hao_new} searched for OCs in the direction of DCEPs and identified 50 DCEPs that may be associated with OCs. While 12 DCEPS may not falling the instability strip of their host OC. Among the 38 OC-DCEP candidates, 28 DCEPs are members of host OCs, and the other 10 DCEPs are not members of OCs, but their five-dimensional parameters (three spatial dimensions and two proper motion dimensions) are within 3$\sigma$ of the five-dimensional parameter space of the associated OCs.

Besides, combining two large known OCs catalogues (2017 OCs catalogued by \citealt{Cantat-Gaudin+2020} and 628 OCs newly certified by \citealt{Castro-Ginard+2022}) with the $\sim$ 3300 DCEP sample of \cite{Pietrukowicz+2021}, we also searched the additional complex of OCs and DCEPs in the five-dimensional parameter space ($\alpha,\delta,\varpi,\mu_{\alpha}\cos \delta,\mu_{\delta}$). The original \cite{Cantat-Gaudin+2020} catalog was based on the astrometric parameters of \emph{Gaia} DR2; here, we have updated this sample using the astrometric parameters from the \emph{Gaia} DR3 catalog. 
Considering the association between the DCEPs and the OCs, their five-dimensional parameters are within 3$\sigma$. However, the known DCEP star clusters No Cas - NGC 103 and V379 Cas - NGC 129 have several dimensions that are slightly higher than 3$\sigma$ but less than 4$\sigma$, which were not considered as OC-DCEPs by the 3$\sigma$ limitation employed here. We also consider an extended sample where several of the dimensions were slightly higher than 3$\sigma$ but less than 4$\sigma$.

We combined the results of \citep{Hao_new}, and cross-matched the OCs of \cite{Cantat-Gaudin+2020,Castro-Ginard+2022} with the DCEPs of \cite{Pietrukowicz+2021}; the total OC-DCEP candidates are shown in Table~\ref{tab1}. We obtained 72 possible candidates.

\begin{figure}[!ht]
  \centering
  \subfigure[]{\includegraphics[width=0.48\textwidth]{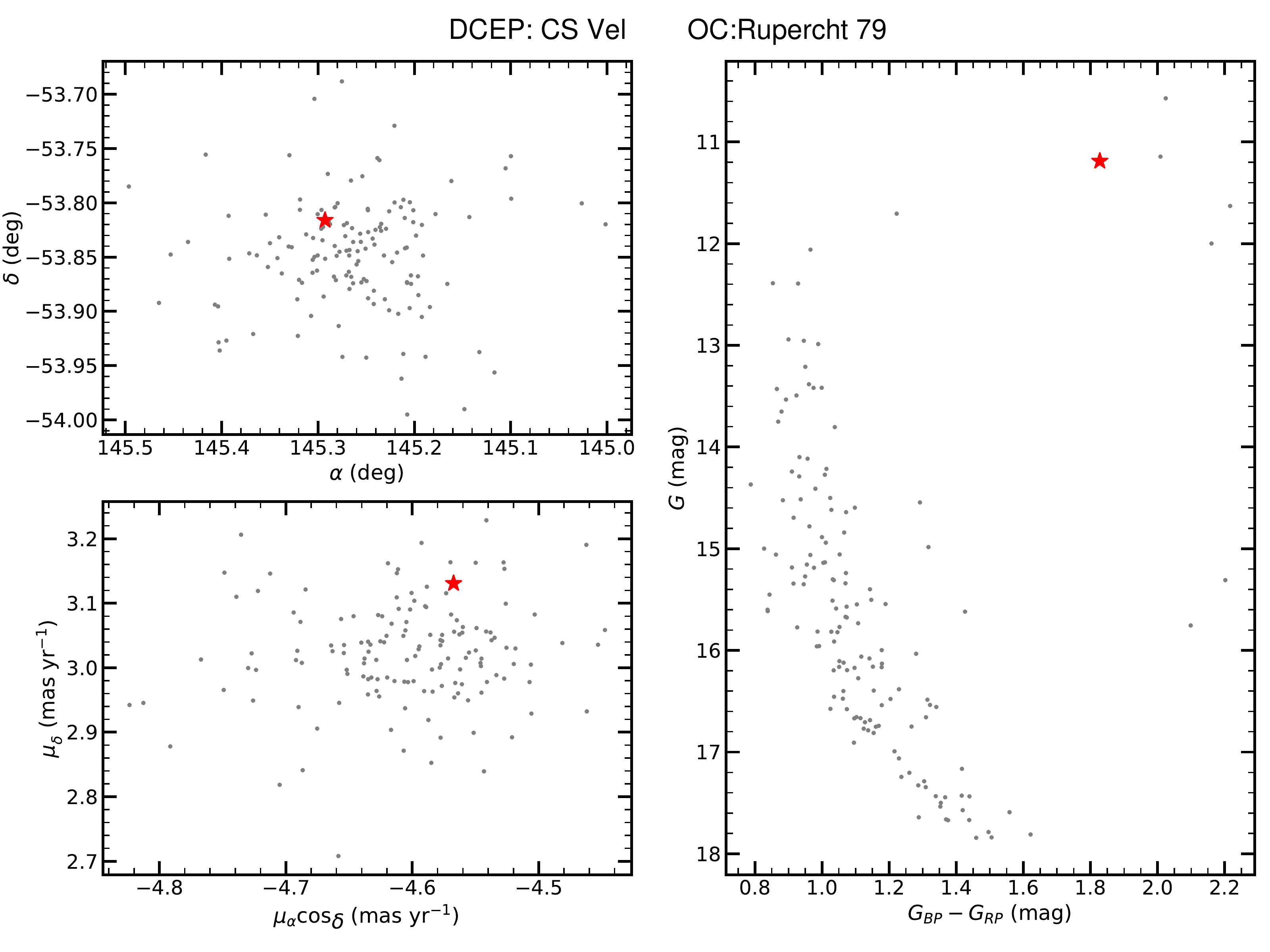}}
  \subfigure[]{\includegraphics[width=0.48\textwidth]{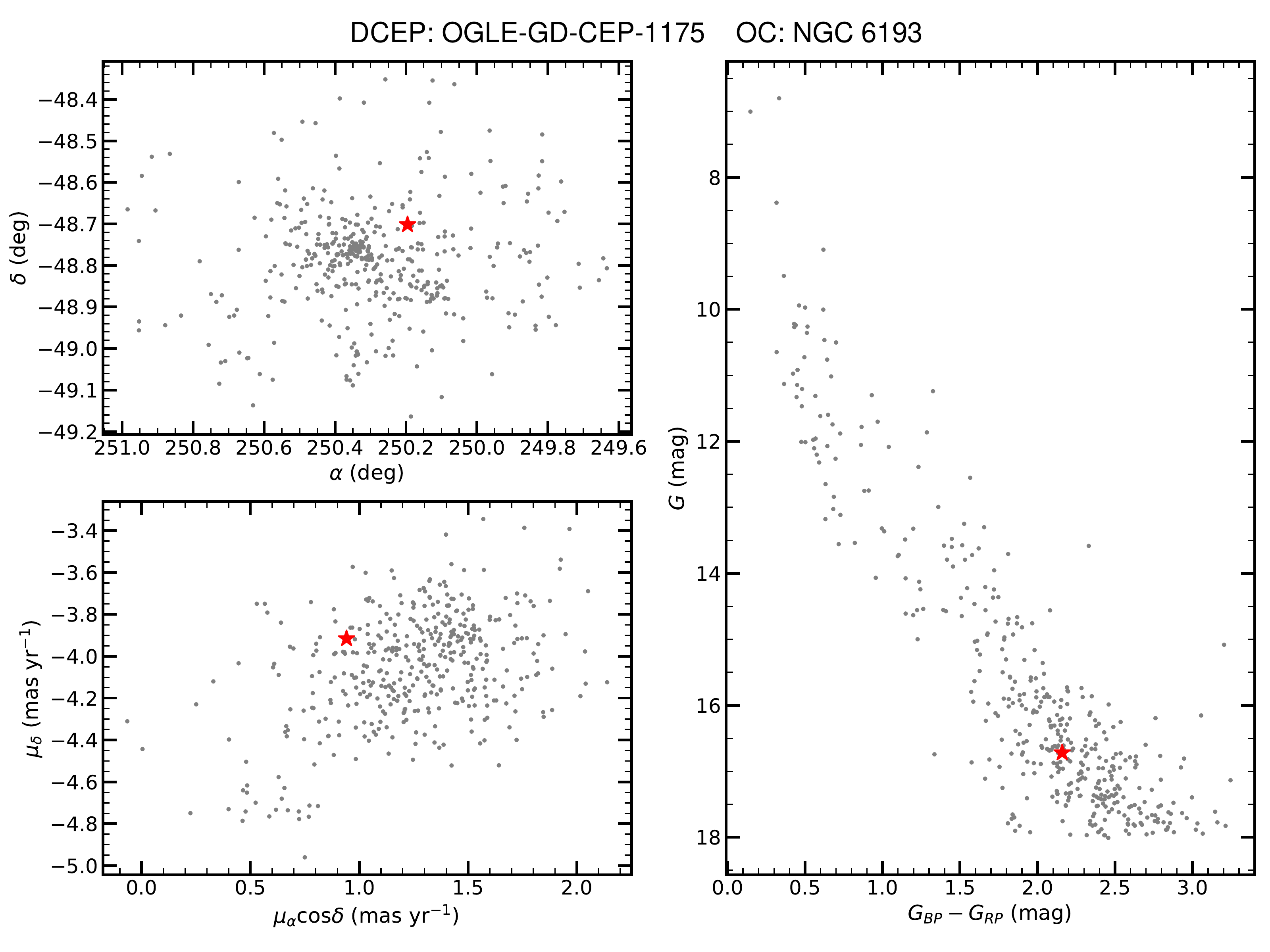}}
  \caption{Examples of classical Cepheid OC combos. (a) Example of the DCEPs CS Vel and the OC Ruprecht 79, showing the spatial distribution (top right), proper-motion distribution (bottom right), and color--magnitude diagram (left). The gray dots are member stars of the OC, and the red asterisks represent the DCEPs. (b) Same distributions for DCEPs OGLE-GD-CEP-1175 and the OC NGC 6193. In the color--magnitude diagram in panel (b), OGLE-GD-CEP-1175 is fainter than the turn-off point of its OC.}
  \label{figA1}
\end{figure}

Figure~\ref{figA1} shows the spatial, proper-motion, and color--magnitude distributions of the OCs and DCEPs. Cepheids, in the stage of helium core burning, should have a magnitude brighter than the turn-off point of the host OC. Thus, we excluded OC-DCEP candidates whose DCEP magnitude was fainter than the OC's turn-off point (as shown in panel (b) of Figure~\ref{figA1}, and marked as “n” in Table~\ref{tab1}). However, OC-DCEP candidates whose number of member stars of the cluster was too few to confirm whether the DCEP was brighter than the turn-off point, were also excluded (marked as “u” in Table~\ref{tab1}). In this step, we rejected 22 samples that were consistent with the OCs in the five-dimensional space, but whose magnitude was fainter than the OC's turn-off point (as listed under the ``unlikely OC-DCEP sample'' in Table~\ref{tab1}).

CG Cas is a member of the OC UBC 406 \citep{Cantat-Gaudin+2020}, while \cite{Hao_new} considered it to be a member of Berkeley 58. 
Considering that the spatial distribution of OC UBC 406 and Berkeley 58 have a very high coincidence, combined with previous OC-DCEP results \citep{Turner+2008}, we also consider that CG Cas is associated with Berkeley 58.

Finally, we obtain 51 OC-DCEPs, including 33 where the DCEP is a member of a host cluster, seven in which the DCEP is associated with a cluster to within 3$\sigma$ in the five-dimensional parameter space, and 10 DCEPs whose five-dimensional space was locate outside 3$\sigma$ but less than 4$\sigma$ of that of an OC.

\section{Open cluster classical Cepheid candidate}\label{secB}

This appendix presents the cross-matched result of the DCEPs and OCs (listed in Table~\ref{tab1}) that are discussed in Appendix~\ref{secA1}, and the parameters of the OC-DCEP candidates (listed in Table~\ref{tab2}) that are discussed in Section~\ref{sec2} and Section~\ref{sec3.1}

\setcounter{table}{0}
\setcounter{figure}{0}
\renewcommand{\thetable}{B\arabic{table}}
\renewcommand{\thefigure}{B\arabic{figure}}

\startlongtable
\begin{deluxetable}{rrrccccccccl}
\setlength\tabcolsep{3pt}
\tablecolumns{12}
\tablecaption{Crossmatched result of the Classical Cepheid and open cluster \label{tab1}}
\tablehead{
\colhead{No.} & \colhead{DCEP.Name} & \colhead{OC.Name}& \colhead{$\Delta_\alpha$} & \colhead{$\Delta_\delta$} & \colhead{$\Delta_\varpi$}  & \colhead{$\Delta_{\mu_\alpha \cos \delta}$} & \colhead{$\Delta_{\mu_\delta}$} & \colhead{$\Delta_{tot}$} & \colhead{ref} & \colhead{members}& \colhead{Note} \\
\colhead{(1)} & \colhead{(2)} & \colhead{(3)} & \colhead{(4)} & \colhead{(5)} & \colhead{(6)} & \colhead{(7)} & \colhead{(8)} & \colhead{(9)} & \colhead{(10)} & \colhead{(11)} & \colhead{(12)}
}
\startdata
 1 & CG Cas & Berkeley 58 & -0.47 &  0.29 & -1.13 &  2.76 &  2.03 &  3.65 & 1,3 & Y &\\
 2 & V824 Cas & UBC 409 & -3.73 & -1.27 & -0.69 &  0.33 & -1.08 &  4.16 & 1 & N & 1e\\
 3 & NO Cas & NGC 103 & -2.51 &  0.42 & -0.27 & -0.35 & -1.80 &  3.15 & 1,3 & Y & \\
 4 & V379 Cas & NGC 129 & -2.40 &  3.97 & -0.76 & -1.07 & -1.32 &  5.00 & 1 & N & 1e\\
 5 & DL Cas & NGC 129 & -0.22 & -0.05 &  0.59 & -1.18 & -0.11 &  1.34 & 1,3 & Y &  \\
 6 & AU Cas & UBC 412 &  2.78 &  3.49 & -1.00 &  0.10 &  3.43 &  5.72 & 1 & N & 2e\\
 7 & SZ Cas & UBC 190 & -3.06 & -2.85 &  0.87 &  3.00 & -1.65 &  5.47 & 1 & N & 2e\\
 8 & SU Cas & Alessi 95 &  0.02 & -0.02 & -1.84 &  4.77 & -0.50 &  5.13 & 3 & Y & \\
 9 & MQ Cam & OC-0715 &  1.08 & -1.87 &  1.11 &  2.47 &  1.13 &  3.64 & 3 & N & \\
10 & AN Aur & UBC1273 & -0.50 &  0.34 &  0.21 & -0.47 & -0.23 &  0.83 & 2 & Y & \\
11 & RS Ori & FSR 0951 & -0.21 &  0.34 & -0.19 & -0.20 & -0.35 &  0.59 & 1,3 & Y & \\
12 & CV Mon & vdBergh 1 & -0.22 & -0.31 &  0.27 & -0.17 &  0.42 &  0.65 & 1,3 & Y & \\
13 & WX Pup & OC-0717 &  0.84 &  0.54 & -0.71 & -0.06 &  0.71 &  1.42 & 3 & Y & \\
14 & V724 Pup & UBC1429 & -1.06 & -0.07 &  1.40 &  1.09 &  1.11 &  2.35 & 2 & Y & \\
15 & V335 Pup & UBC 229 & -0.37 & -0.24 &  1.11 &  0.22 &  0.27 &  1.25 & 1,3 & Y & \\
16 & J075840-3330.2 & UBC1424 &  -0.43 &  1.20 &  0.58 & 0.05 &  0.88 &  1.66 & 3,2 & Y & \\
17 & J084951-4627.2 & OC-0723 &  1.18 &  0.95 &  1.15 & -2.68 & -0.95 &  3.42 & 3 & N & \\
18 & DP Vel & UBC 491 & -1.68 & -0.52 &  1.56 &  0.42 &  0.24 &  2.39 & 1,3 & N &\\
19 & CS Vel & Ruprecht 79 &  0.37 &  0.59 &  0.29 &  0.54 &  1.34 &  1.63 & 1,3 & Y & \\
20 & DY Car & UBC1490 & -3.28 &  1.27 &  0.03 &  3.50 &  2.23 &  5.44 & 2 & N & 2e\\
21 & WZ Car & UBC1496 & -2.16 &  3.47 &  2.70 &  2.71 &  1.29 &  5.75 & 2 & N & 1e\\
22 & X Cru & UBC 290 & -0.96 &  2.28 &  1.21 &  0.07 &  0.26 &  2.77 & 1,3 & N & \\
23 & V Cen & NGC 5662 & -1.36 & -0.98 &  1.57 & -1.46 &  0.72 &  2.81 & 1,3 & Y & \\
24 & OGLE-GD-CEP-1012 & Teutsch 80 &  3.29 &  0.71 & -2.79 &  2.67 &  1.06 &  5.23 & 1 & N & 1e\\
25 & TW Nor & Lynga 6 &  0.43 &  0.09 & -1.61 & -0.25 & -0.09 &  1.70 & 1,3 & N & \\
26 & V340 Nor & NGC 6067 &  0.18 & -0.12 & -0.17 & -0.22 & -0.42 &  0.88 & 1,3 & Y & \\
27 & S Nor& NGC 6087 & -0.28 &  -0.23 &  1.57 &  0.12 &  1.44 &  2.16 & 1,3 & Y & \\
28 & KQ Sco & UBC1558 &  1.64 & -0.28 &  3.40 & -0.56 &  0.27 &  3.84 & 2 & N & 1e\\
29 & VdBH222-505 & BH 222 &  0.23 & -0.14 &  0.38 & -0.55 & -1.05 &  1.27 & 1,3 & Y & \\
30 & U Sgr & IC 4725 &  0.15 & -0.06 &  0.75 & -0.49 &  0.25 &  0.95 & 1,3 & Y & \\
31 & V367 Sct & NGC 6649 &  0.74 & -0.54 & -0.62 &  0.39 & -1.04 &  1.57 & 1,3 & Y & \\
32 & EV Sct & NGC 6664 &  0.68 &  -0.08 &  0.72 & -0.99 &  0.37 &  1.45 & 1,3 & Y & \\
33 & CM Sct & UBC 106 &  1.13 &  0.60 &  0.06 & -0.16 & -0.41 &  1.35 & 1,3 & Y & \\
34 & CN Sct & Trumpler 35 & -1.99 & -1.93 &  0.46 & -0.55 & -0.24 &  2.87 & 1,3 & N & \\
35 & J280.6497-05.2660 & Teutsch 145 &  2.05 & -1.35 & -1.74 & -1.21 &  3.54 &  4.80 & 1 & N & 1e\\
36 & GQ Vul & FSR 0158 &  -0.24 & -0.71 & -0.48 &  -0.42 & 0.31 &  1.03 & 1,3 & Y & \\
37 & J194806.54+260526.1 & FSR 0158 &  0.44 & 1.13 & 3.74 &  -1.29 & -0.44 & 4.16 & 1,3 & N & 1e\\
38 & J297.7863+25.3136 & Czernik 41 &  0.68 &  0.48 & -1.02 & -0.76 & -1.06 &  1.85 & 1,3 & Y & \\
39 & SV Vul & UBC 130 & -1.48 &  0.19 & -0.54 & -0.40 & -0.51 &  1.72 & 1,3 & Y & \\
40 & X Vul & UBC 129 &  1.73 &  0.50 & -0.62 & -2.40 & 0.53 &  3.11 & 3 & Y & \\
41 & GI Cyg & UBC 135 &  1.24 &  0.31 &  0.16 &  0.84 & -1.32 &  2.02 & 1,3 & Y & \\
42 & CD Cyg & UBC1089 & -1.00 &  2.36 &  0.70 &  0.37 & -0.27 &  2.69 & 2 & Y & \\
43 & J201151.18+342447.2 & Berkeley 51 & -0.37 &  0.78 &  0.53 & -1.01 &  0.12 &  1.44 & 1,3 & Y & \\
44 & V1788 Cyg & OC-0125 & -1.28 &  0.13 &  0.15 &  0.14 & -1.00 &  1.64 & 3 & Y & \\
45 & J211659.94+514556.7 & Berkeley 55 &  0.21 & -0.03 &  0.25 &  0.95 & -0.45 &  1.11 & 1,3 & Y & \\
46 & V733 Cyg & Kronberger 84 & -2.26 & -2.08 &  0.96 &  1.26 & -2.03 &  4.01 & 3 & N & \\
47 & J213533.70+533049.3 & Kronberger 84 &  -0.11 & -0.16 & -0.07 &  -0.27 & -0.54 &  0.64 & 1,3 & Y & \\
48 & J233736.95+602243.8 & SAI 149 & -0.77 & -3.30 &  0.59 &  2.83 & -0.16 &  4.46 & 1 & N & 1e\\
49 & CE Cas B & NGC 7790 & -0.59 &  0.02 &  0.34 & -0.67 & -1.20 &  1.54 & 1,3 & Y & \\
50 & CE Cas A & NGC 7790 & -0.58 &  0.02 &  0.31 & -0.63 & -2.08 &  2.27 & 1,3 & Y & \\
51 & CF Cas & NGC 7790 & -0.32 &  0.17 & -0.10 &  0.11 & -0.61 &  0.73 & 1,3 & Y & \\
\hline
\multicolumn{12}{c}{unlikely OC-DCEP samples}\\
\hline
 & J040516.13+555512.9 & UBC 608 & -0.42 & -0.14 & -0.50 & -0.36 &  2.07 &  2.21 & 1 & Y & n\\
 & OGLE-GD-CEP-0059 & UBC1339 &  0.85 &  2.55 & -3.12 &  0.01 & -3.55 &  5.44 & 2 & N & n,2e\\
 & OGLE-GD-CEP-0270 & IC 2395 &  1.52 &  1.85 & -0.23 & -1.54 &  0.10 &  2.85 & 1 & Y & n\\
 & OGLE-GD-CEP-1631 & Ruprecht 63 &  2.93 & -3.61 & -3.08 &  2.22 & -3.38 &  6.89 & 1 & N & n,3e\\
 & OGLE-GD-CEP-0436 & Schuster 1 & -1.96 &  0.54 &  1.20 & -0.65 & -0.90 &  2.61 & 1 & N & n\\
 & V701 Car & FSR 1530 &  1.41 &  1.66 &  2.36 &  0.77 & -2.55 &  4.17 & 1 & N & n\\
 & OGLE-GD-CEP-0549 & UBC 261 & -2.41 & -2.09 & -2.50 &  1.57 &  1.79 &  4.70 & 1 & N & n\\
 & OGLE-GD-CEP-1673 & Trumpler 14 &  0.07 & -0.69 & -1.23 & -0.14 &  0.08 &  1.42 & 1 & Y & n\\
 & OGLE-GD-CEP-0646 & NGC 3603 &  3.76 &  1.54 & -0.55 & -2.44 & -1.31 &  4.95 & 1 & N & n,1e\\
 & OGLE-GD-CEP-1688 & BH 121 &  0.73 & -1.10 &  0.21 &  0.86 & -1.67 &  2.30 & 1 & Y & n\\
 & D19-158 & UPK 604 & -1.37 &  1.32 & -2.51 & -0.57 &  0.89 &  3.32 & 1 & N & n\\
 & OGLE-GD-CEP-1752 & ASCC 79 &  0.95 & -0.61 & -1.90 &  0.48 & -0.28 &  2.28 & 1 & N & n\\
 & OGLE-GD-CEP-1172 & UBC1550 &  3.35 & -0.65 & -2.86 & -0.76 &  3.60 &  5.77 & 2 & N & n,2e\\
 & OGLE-GD-CEP-1175 & NGC 6193 & -0.42 &  0.66 & -0.49 & -0.85 &  0.54 &  1.37 & 1 & Y & n\\
 & OGLE-GD-CEP-1194 & UBC 553 & -0.69 & -0.23 & -2.01 &  0.18 &  0.60 &  2.23 & 1 & N & n\\
 & OGLE-GD-CEP-1196 & UBC 323 & -1.12 & -2.88 &  3.55 & -1.77 & -2.28 &  5.52 & 1 & N & n,1e\\
 & OGLE-BLG-CEP-172 & Gulliver 29 &  1.36 &  1.65 &  1.10 &  1.53 &  0.07 &  2.85 & 1 & N & n\\
 & OGLE-BLG-CEP-157 & NGC 6611 &  2.81 & -3.39 & -2.63 &  0.10 &  1.78 &  5.43 & 1 & N & 1e,n\\
 & OGLE-BLG-CEP-164 & NGC 6631 & -1.32 &  0.44 & -3.93 & -0.47 &  2.88 &  5.09 & 1 & N & 1e,n\\
 & J300.0102+29.1869 & FSR 0172 &  0.25 & -1.35 &  1.00 &  0.44 & -1.68 &  2.43 & 1 & N & n\\
 & J201359.38+361435.2 & UBC 583 & -1.60 & -3.44 & -3.43 &  1.24 &  0.39 &  5.28 & 1 & N & n,2e\\
 & J211037.61+483520.5 & Berkeley 91 & -1.09 &  1.53 & -3.67 & -3.17 &  0.92 &  5.28 & 1 & N & u,2e\\
\enddata
\vspace{0.5cm}
Note--
\begin{itemize}
  \item [--] Column (1): The index of the OC-DCEP candidate. This work is only used first 51 rows and the following samples without index is unlikely OC-DCEP candidate with the check in color-magnitude diagram.
  \item [--] Column (2): The name of the classic Cepheid.
  \item [--] Column (3): The name of the cluster that may be associated.
  \item [--] Columns (4)-(8): The deviation between the classical Cepheid and the open cluster in the five-dimensional parameters ($\alpha$, $\delta$, $\varpi$, $\mu_\alpha \cos \delta$, $\mu_\delta$). Its value is expressed as several times of the standard deviation. If this OC-DCEP candidate identified with two reference, we only exhibit the results that total deviation is smallest.
  \item [--] Column (9): The total deviation as $\Delta_{tot} = \sqrt{\Delta_\alpha^{2} + \Delta_\delta^2 + \Delta_\varpi^2 + \Delta_{\mu_\alpha \cos \delta}^2 + \Delta_{\mu_\delta}^2}$.
  \item [--] Column (10): The reference of the open cluster, where 1 is the catalogue of \cite{Cantat-Gaudin+2020}, 2 is the catalogue of \cite{Castro-Ginard+2022} and 3 is the catalogue of \cite{Hao_new}.
  \item [--] Column (11): The classical Cepheid is the member of open cluster or not.
  \item [--] Column (12): u means that the number of member stars of open cluster is few, so it is difficult to judge whether Cepheid is above the turn-off point. n means that Cepheid is below the turn-off point of the cluster, and e means that some dimensions are outside 3$\sigma$. For example, 2e is a classic Cepheid with two dimension outside 3$\sigma$ of the cluster.
\end{itemize}
\end{deluxetable}


\clearpage
\startlongtable
\begin{deluxetable}{ccccccccccccc}
\setlength\tabcolsep{3pt}
\tablecolumns{13}
\tablecaption{Parameters of the open cluster classical Cepheid candidates\label{tab2}}
\tablehead{
 \colhead{No.} & \colhead{mode} & \colhead{P} & \colhead{$m_G$} & \colhead{$\sigma_{m_G}$} & \colhead{$m_{G_{BP}} - m_{G_{RP}}$} & \colhead{$\sigma_{(m_{G_{BP}} - m_{G_{RP}})}$} & \colhead{$\varpi_{DCEP}$} & \colhead{$\sigma_{\varpi_{DCEP}}$} & \colhead{$\varpi_{OC}$} & \colhead{$\sigma_{\varpi_{OC}}$} & \colhead{$DM$} & \colhead{$\sigma_{DM}$}
}
\startdata
 1 &    F &  4.37 & 10.843 & 0.003 &  1.643 & 0.015 & 0.268 & 0.014 & 0.308 & 0.029 & 12.70 & 0.2 \\
 2 &   1O &  5.35 & 10.590 & 0.001 &  1.777 & 0.004 & 0.266 & 0.013 & 0.308 & 0.026 & 12.41 & 0.2 \\
 3 &   1O &  2.58 & 10.973 & 0.001 &  1.489 & 0.010 & 0.270 & 0.013 & 0.289 & 0.034 & 12.50 & 0.2 \\
 4 &   1O &  4.31 &  8.690 & 0.002 &  1.465 & 0.003 & 0.498 & 0.014 & 0.527 & 0.033 & 11.40 & 0.1 \\
 5 &    F &  8.00 &  8.515 & 0.001 &  1.519 & 0.003 & 0.553 & 0.027 & 0.527 & 0.033 & 11.40 & 0.1 \\
 6 &    F &  5.62 & 12.185 & 0.002 &  2.202 & 0.002 & 0.202 & 0.013 & 0.238 & 0.034 & 12.71 & 0.2 \\
 7 &    F & 13.64 &  9.141 & 0.001 &  1.916 & 0.001 & 0.373 & 0.017 & 0.348 & 0.022 & 12.54 & 0.2 \\
 8 &   1O &  1.95 &  5.736 & 0.001 &  1.001 & 0.006 & 2.166 & 0.062 & 2.275 & 0.060 &  8.12 & 0.1 \\
 9 &    F &  6.60 & 11.929 & 0.001 &  2.236 & 0.005 & 0.248 & 0.018 & 0.233 & 0.015 & 13.05 & 0.2 \\
10 &    F & 10.29 &  9.949 & 0.018 &  1.566 & 0.022 & 0.249 & 0.018 & 0.247 & 0.015 & 13.23 & 0.2 \\
11 &    F &  7.57 &  8.093 & 0.012 &  1.313 & 0.017 & 0.555 & 0.030 & 0.568 & 0.039 & 11.22 & 0.1 \\
12 &    F &  5.38 &  9.659 & 0.003 &  1.802 & 0.003 & 0.567 & 0.015 & 0.558 & 0.045 & 11.45 & 0.2 \\
13 &    F &  8.94 &  8.734 & 0.001 &  1.255 & 0.002 & 0.368 & 0.015 & 0.371 & 0.004 & 11.95 & 0.2 \\
14 &    F &  5.56 & 10.572 & 0.003 &  1.669 & 0.005 & 0.298 & 0.011 & 0.279 & 0.015 & 12.87 & 0.2 \\
15 &   1O &  4.86 &  8.467 & 0.001 &  1.047 & 0.003 & 0.424 & 0.018 & 0.391 & 0.026 & 12.04 & 0.2 \\
16 &    F &  4.40 & 11.634 & 0.001 &  1.654 & 0.001 & 0.242 & 0.013 & 0.236 & 0.015 & 13.22 & 0.2 \\
17 &   1O &  3.79 & 10.819 & 0.001 &  1.493 & 0.001 & 0.215 & 0.014 & 0.183 & 0.016 & 13.40 & 0.2 \\
18 &    F &  5.48 & 11.183 & 0.001 &  1.837 & 0.002 & 0.325 & 0.013 & 0.275 & 0.022 & 12.87 & 0.2 \\
19 &    F &  5.90 & 11.098 & 0.001 &  1.782 & 0.003 & 0.261 & 0.012 & 0.249 & 0.025 & 12.90 & 0.2 \\
20 &    F &  4.67 & 10.944 & 0.002 &  1.344 & 0.005 & 0.205 & 0.013 & 0.206 & 0.018 & 13.58 & 0.2 \\
21 &    F & 23.02 &  8.830 & 0.013 &  1.454 & 0.017 & 0.271 & 0.018 & 0.250 & 0.007 & 12.91 & 0.2 \\
22 &    F &  6.22 &  8.080 & 0.002 &  1.251 & 0.004 & 0.638 & 0.019 & 0.609 & 0.031 & 11.03 & 0.1 \\
23 &    F &  5.49 &  6.534 & 0.002 &  1.167 & 0.007 & 1.390 & 0.022 & 1.298 & 0.042 &  9.41 & 0.1 \\
24 &    F & 15.97 & 12.086 & 0.005 &  3.170 & 0.009 & 0.177 & 0.027 & 0.351 & 0.033 & 11.93 & 0.2 \\
25 &    F & 10.79 & 10.526 & 0.007 &  2.536 & 0.011 & 0.318 & 0.020 & 0.386 & 0.032 & 11.92 & 0.2 \\
26 &    F & 11.29 &  7.993 & 0.001 &  1.405 & 0.003 & 0.466 & 0.025 & 0.473 & 0.038 & 11.37 & 0.1 \\
27 &    F &  9.75 &  6.165 & 0.004 &  1.163 & 0.007 & 1.077 & 0.022 & 1.028 & 0.056 &  9.90 & 0.1 \\
28 &    F & 28.70 &  8.853 & 0.005 &  2.318 & 0.017 & 0.432 & 0.021 & 0.386 & 0.023 & 12.44 & 0.2 \\
29 &    F & 23.30 & 13.014 & 0.023 &  3.720 & 0.016 & 0.166 & 0.036 &   \nodata &   \nodata & 13.65 & 0.2 \\
30 &    F &  6.75 &    \nodata &   \nodata &    \nodata &   \nodata & 1.569 & 0.022 & 1.517 & 0.064 &  9.22 & 0.1 \\
31 &  F1O &  6.29 & 10.508 & 0.011 &  2.417 & 0.016 & 0.425 & 0.020 & 0.477 & 0.041 & 11.64 & 0.1 \\
32 &   1O &  3.09 &  9.624 & 0.001 &  1.605 & 0.005 & 0.489 & 0.018 & 0.465 & 0.038 & 11.62 & 0.1 \\
33 &    F &  3.92 & 10.515 & 0.003 &  1.806 & 0.007 & 0.405 & 0.016 & 0.406 & 0.028 & 11.86 & 0.1 \\
34 &    F &  9.99 & 11.142 & 0.004 &  2.779 & 0.007 & 0.359 & 0.030 & 0.345 & 0.031 & 12.22 & 0.2 \\
35 &    F & 14.36 & 12.825 & 0.024 &  3.937 & 0.022 & 0.084 & 0.054 &   \nodata &   \nodata & 12.75 & 0.2 \\
36 &    F & 12.66 & 12.221 & 0.006 &  2.934 & 0.020 & 0.137 & 0.020 &   \nodata &   \nodata & 13.94 & 0.2 \\
37 &   1O & 6.65 & 12.399 & 0.001 &  2.858 & 0.002 & 0.170 & 0.018 &   \nodata &   \nodata & 13.94 & 0.2 \\
38 &   1O &  2.94 & 12.096 & 0.001 &  2.971 & 0.003 & 0.325 & 0.018 & 0.373 & 0.036 & 12.03 & 0.1 \\
39 &    F & 44.89 &  6.665 & 0.005 &  1.703 & 0.007 & 0.373 & 0.021 & 0.398 & 0.023 & 11.82 & 0.2 \\
40 &    F &  6.32 &  8.238 & 0.002 &  1.827 & 0.003 & 0.836 & 0.022 & 0.852 & 0.024 & 10.41 & 0.1 \\
41 &    F &  5.78 & 11.066 & 0.002 &  1.862 & 0.006 & 0.247 & 0.014 & 0.241 & 0.022 & 13.15 & 0.2 \\
42 &    F & 17.08 &  8.456 & 0.010 &  1.624 & 0.020 & 0.367 & 0.016 & 0.357 & 0.015 & 12.40 & 0.2 \\
43 &    F &  9.83 & 13.565 & 0.009 &  3.144 & 0.012 & 0.186 & 0.014 & 0.163 & 0.016 & 13.60 & 0.2 \\
44 &    F & 14.09 & 11.189 & 0.025 &  2.902 & 0.019 & 0.348 & 0.021 & 0.345 & 0.032 & 12.33 & 0.2 \\
45 &    F &  5.85 & 12.369 & 0.001 &  2.865 & 0.003 & 0.324 & 0.018 & 0.323 & 0.028 & 12.20 & 0.2 \\
46 &    F &  4.56 & 11.964 & 0.001 &  2.051 & 0.006 & 0.223 & 0.012 & 0.201 & 0.025 & 13.03 & 0.2 \\
47 &   1O &  3.20 & 11.897 & 0.001 &  1.766 & 0.002 & 0.193 & 0.010 & 0.201 & 0.025 & 13.03 & 0.2 \\
48 &    F &  3.90 & 12.917 & 0.004 &  2.643 & 0.005 & 0.362 & 0.018 & 0.334 & 0.071 & 12.67 & 0.2 \\
49 &    F &  4.48 & 10.585 & 0.002 &  1.471 & 0.042 & 0.307 & 0.015 & 0.306 & 0.032 & 12.60 & 0.2 \\
50 &    F &  5.14 & 10.498 & 0.001 &  1.494 & 0.038 & 0.306 & 0.015 & 0.306 & 0.032 & 12.60 & 0.2 \\
51 &    F &  4.88 & 10.658 & 0.001 &  1.570 & 0.005 & 0.289 & 0.012 & 0.306 & 0.032 & 12.60 & 0.2 \\
\enddata
\tablecomments{The parameters of 51 open cluster classical Cepheid candidates, including pulsation mode (mode), period (P), intensity-averaged magnitude in $G$ band ($m_G$), uncertainty on intensity-averaged magnitude in $G$ band ($\sigma_{m_G}$), colour excess ($m_{G_{BP}} - m_{G_{RP}}$), uncertainty of color excess ($\sigma_{(m_{G_{BP}} - m_{G_{RP}})}$), parallax of classical Cepheid ($\varpi_{DCEP}$), standard error of classical Cepheid parallax ($\sigma_{\varpi_{DCEP}}$), parallax of host open cluster ($\varpi_{OC}$), standard deviation of host open cluster parallax ($\sigma_{\varpi_{OC}}$), distance modulus ($DM$), and calculated uncertainty of distance modulus ($\sigma_{DM}$).}
\end{deluxetable}

\end{appendix}


\begin{thebibliography}{aa}
\bibitem[Anderson et al.(2013)]{Anderson+2013} Anderson, R.~I., Eyer, L., \& Mowlavi, N.\ 2013, \mnras, 434, 2238. doi:10.1093/mnras/stt1160
\bibitem[Andrae et al.(2018)]{Andrae+2018} Andrae, R., Fouesneau, M., Creevey, O., et al.\ 2018, \aap, 616, A8. doi:10.1051/0004-6361/201732516
\bibitem[Breuval et al.(2020)]{Breuval+2020} Breuval, L., Kervella, P., Anderson, R.~I., et al.\ 2020, \aap, 643, A115. doi:10.1051/0004-6361/202038633
\bibitem[Cantat-Gaudin et al.(2018)]{Cantat-Gaudin+2018} Cantat-Gaudin, T., Jordi, C., Vallenari, A., et al.\ 2018, \aap, 618, A93. doi:10.1051/0004-6361/201833476
\bibitem[Cantat-Gaudin et al.(2019)]{Cantat-Gaudin+2019} Cantat-Gaudin, T., Krone-Martins, A., Sedaghat, N., et al.\ 2019, \aap, 624, A126. doi:10.1051/0004-6361/201834453
\bibitem[Cantat-Gaudin et al.(2020)]{Cantat-Gaudin+2020} Cantat-Gaudin, T., Anders, F., Castro-Ginard, A., et al.\ 2020, \aap, 640, A1. doi:10.1051/0004-6361/202038192
\bibitem[Cardelli et al.(1989)]{Cardelli+1989} Cardelli, J.~A., Clayton, G.~C., \& Mathis, J.~S.\ 1989, \apj, 345, 245. doi:10.1086/167900
\bibitem[Castro-Ginard et al.(2018)]{Castro-Ginard+2018} Castro-Ginard, A., Jordi, C., Luri, X., et al.\ 2018, \aap, 618, A59. doi:10.1051/0004-6361/201833390
\bibitem[Castro-Ginard et al.(2019)]{Castro-Ginard+2019} Castro-Ginard, A., Jordi, C., Luri, X., et al.\ 2019, \aap, 627, A35. doi:10.1051/0004-6361/201935531
\bibitem[Castro-Ginard et al.(2020)]{Castro-Ginard+2020} Castro-Ginard, A., Jordi, C., Luri, X., et al.\ 2020, \aap, 635, A45. doi:10.1051/0004-6361/201937386
\bibitem[Castro-Ginard et al.(2022)]{Castro-Ginard+2022} Castro-Ginard, A., Jordi, C., Luri, X., et al.\ 2022, \aap, 661, A118. doi:10.1051/0004-6361/202142568
\bibitem[Chen et al.(2019)]{Chen+2019} Chen, X., Wang, S., Deng, L., et al.\ 2019, Nature Astronomy, 3, 320. doi:10.1038/s41550-018-0686-7
\bibitem[Dias et al.(2021)]{Dias+2021} Dias, W.~S., Monteiro, H., Moitinho, A., et al.\ 2021, \mnras, 504, 356. doi:10.1093/mnras/stab770
\bibitem[ESA(1997)]{ESA+1997} ESA\ 1997, ESA Special Publication, 1200
\bibitem[Fabricius et al.(2021)]{Fabricius+2021} Fabricius, C., Luri, X., Arenou, F., et al.\ 2021, \aap, 649, A5. doi:10.1051/0004-6361/202039834
\bibitem[Feast \& Catchpole(1997)]{Feast_Catchpole_1997} Feast, M.~W. \& Catchpole, R.~M.\ 1997, \mnras, 286, L1. doi:10.1093/mnras/286.1.L1
\bibitem[Fitzpatrick et al.(2019)]{Fitzpatrick+2019} Fitzpatrick, E.~L., Massa, D., Gordon, K.~D., et al.\ 2019, \apj, 886, 108. doi:10.3847/1538-4357/ab4c3a
\bibitem[Gaia Collaboration et al.(2016)]{Gaia+2016} Gaia Collaboration, Prusti, T., de Bruijne, J.~H.~J., et al.\ 2016, \aap, 595, A1. doi:10.1051/0004-6361/201629272
\bibitem[Gaia Collaboration et al.(2018)]{Gaiadr2+2018} Gaia Collaboration, Babusiaux, C., van Leeuwen, F., et al.\ 2018, \aap, 616, A10. doi:10.1051/0004-6361/201832843
\bibitem[Gaia Collaboration et al.(2021)]{Gaiaedr3+2021} Gaia Collaboration, Brown, A.~G.~A., Vallenari, A., et al.\ 2021, \aap, 649, A1. doi:10.1051/0004-6361/202039657
\bibitem[Gaia Collaboration et al.(2022)]{Gaiadr3+2022} Gaia Collaboration, Vallenari, A. Brown, A.~G.~A., et al.\ 2022, \aap, doi:10.1051/0004-6361/202243940
\bibitem[Gordon et al.(2003)]{Gordon+2003} Gordon, K.~D., Clayton, G.~C., Misselt, K.~A., et al.\ 2003, \apj, 594, 279. doi:10.1086/376774
\bibitem[Hao et al.(2020)]{Hao+2020} Hao, C., Xu, Y., Wu, Z., et al.\ 2020, \pasp, 132, 034502. doi:10.1088/1538-3873/ab694d
\bibitem[Hao et al.(2022)]{Hao+2022} Hao, C.~J., Xu, Y., Wu, Z.~Y., et al.\ 2022, \aap, 660, A4. doi:10.1051/0004-6361/202243091
\bibitem[Hao et al.(2022b)]{Hao_new} Hao, C.~J., Xu, Y., Wu, Z., et al.\ 2022, arXiv:2210.01521.
\bibitem[Heinze et al.(2018)]{Heinze+2018} Heinze, A.~N., Tonry, J.~L., Denneau, L., et al.\ 2018, \aj, 156, 241. doi:10.3847/1538-3881/aae47f
\bibitem[Irwin(1955)]{Irwin_1955} Irwin, J.~B.\ 1955, Monthly Notes of the Astronomical Society of South Africa, 14, 38
\bibitem[Jayasinghe et al.(2020)]{Jayasinghe+2020} Jayasinghe, T., Stanek, K.~Z., Kochanek, C.~S., et al.\ 2020, \mnras, 491, 13. doi:10.1093/mnras/stz2711
\bibitem[Leavitt \& Pickering(1912)]{Leavitt_Pickering_1912} Leavitt, H.~S. \& Pickering, E.~C.\ 1912, Harvard College Observatory Circular, 173
\bibitem[Lindegren et al.(2021)]{Lindegren+2021} Lindegren, L., Bastian, U., Biermann, M., et al.\ 2021, \aap, 649, A4. doi:10.1051/0004-6361/202039653
\bibitem[Madore(1982)]{Madore_1982} Madore, B.~F.\ 1982, \apj, 253, 575. doi:10.1086/159659
\bibitem[Madore et al.(2017)]{Madore+2017} Madore, B.~F., Freedman, W.~L., \& Moak, S.\ 2017, \apj, 842, 42. doi:10.3847/1538-4357/aa6e4d
\bibitem[Majaess et al.(2008)]{Majaess+2008} Majaess, D.~J., Turner, D.~G., \& Lane, D.~J.\ 2008, \mnras, 390, 1539. doi:10.1111/j.1365-2966.2008.13834.x
\bibitem[Medina et al.(2021)]{Medina+2021} Medina, G.~E., Lemasle, B., \& Grebel, E.~K.\ 2021, \mnras, 505, 1342. doi:10.1093/mnras/stab1267
\bibitem[Owens et al.(2022)]{Owens+2022} Owens, K.~A., Freedman, W.~L., Madore, B.~F., et al.\ 2022, \apj, 927, 8. doi:10.3847/1538-4357/ac479e
\bibitem[Pietrukowicz et al.(2021)]{Pietrukowicz+2021} Pietrukowicz, P., Soszy{\'n}ski, I., \& Udalski, A.\ 2021, \actaa, 71, 205. doi:10.32023/0001-5237/71.3.2
\bibitem[Pojmanski et al.(2005)]{Pojmanski+2005} Pojmanski, G., Pilecki, B., \& Szczygiel, D.\ 2005, \actaa, 55, 275
\bibitem[Riess et al.(2018)]{Riess+2018} Riess, A.~G., Casertano, S., Yuan, W., et al.\ 2018, \apj, 855, 136. doi:10.3847/1538-4357/aaadb7
\bibitem[Riess et al.(2021)]{Riess+2021} Riess, A.~G., Casertano, S., Yuan, W., et al.\ 2021, \apjl, 908, L6. doi:10.3847/2041-8213/abdbaf
\bibitem[Riess et al.(2022)]{Riess+2022} Riess, A.~G., Breuval, L., Yuan, W., et al.\ 2022, arXiv:2208.01045
\bibitem[Ripepi et al.(2019)]{Ripepi+2019} Ripepi, V., Molinaro, R., Musella, I., et al.\ 2019, \aap, 625, A14. doi:10.1051/0004-6361/201834506
\bibitem[Ripepi et al.(2022a)]{Ripepi+2022} Ripepi, V., Catanzaro, G., Clementini, G., et al.\ 2022, \aap, 659, A167. doi:10.1051/0004-6361/202142649
\bibitem[Ripepi et al.(2022b)]{Ripepi+2022b} Ripepi, V., Clementini, G., Molinaro, R., et al.\ 2022, arXiv:2206.06212
\bibitem[Sandage \& Tammann(2006)]{Sandage_Tammann_2006} Sandage, A. \& Tammann, G.~A.\ 2006, \araa, 44, 93. doi:10.1146/annurev.astro.43.072103.150612
\bibitem[Skowron et al.(2019)]{Skowron+2019} Skowron, D.~M., Skowron, J., Mr{\'o}z, P., et al.\ 2019, Science, 365, 478. doi:10.1126/science.aau3181
\bibitem[Soszy{\'n}ski et al.(2020)]{Soszynski+2020} Soszy{\'n}ski, I., Udalski, A., Szyma{\'n}ski, M.~K., et al.\ 2020, \actaa, 70, 101. doi:10.32023/0001-5237/70.2.2
\bibitem[Turner et al.(2008)]{Turner+2008} Turner, D.~G., Forbes, D., English, D., et al.\ 2008, \mnras, 388, 444. doi:10.1111/j.1365-2966.2008.13413.x
\bibitem[Udalski et al.(2018)]{Udalski+2018} Udalski, A., Soszy{\'n}ski, I., Pietrukowicz, P., et al.\ 2018, \actaa, 68, 315. doi:10.32023/0001-5237/68.4.1
\bibitem[Wang et al.(2017)]{Wang+2017} Wang, S., Jiang, B.~W., Zhao, H., et al.\ 2017, \apj, 848, 106. doi:10.3847/1538-4357/aa8db7
\bibitem[Wang et al.(2018)]{Wang+2018} Wang, S., Chen, X., de Grijs, R., et al.\ 2018, \apj, 852, 78. doi:10.3847/1538-4357/aa9d99
\bibitem[Wang \& Chen(2019)]{Wang_Chen_2019} Wang, S. \& Chen, X.\ 2019, \apj, 877, 116. doi:10.3847/1538-4357/ab1c61
\bibitem[Zhou \& Chen(2021)]{Zhou_Chen_2021} Zhou, X. \& Chen, X.\ 2021, \mnras, 504, 4768. doi:10.1093/mnras/stab1209
\end{thebibliography}
\end{document}